\documentclass[preprint]{aastex631}

\usepackage{url}
\usepackage{mathtools}
\let\tablenum\relax
\usepackage{siunitx}
\newcommand{\J}{CHIME J0630+25}

\newcommand{\meerkat}[0]{PSR J0901--4046}

\newcommand\ILT{ILT J1101+5521}
\newcommand\GLEAM{GLEAM-X J0704-37}
\newcommand\ARSCO{AR Scorpii}
\newcommand\pelisoliWD{J191213.72--441045.1}

\newcommand\gleamtwo{GLEAM-X J0704-37}
\DeclarePairedDelimiterXPP{\BigOSI}[2]%
{\mathcal{O}}{(}{)}{}%
{\SI{#1}{#2}}
\newcommand{\red}[1]{\textcolor{black}{#1}}
\newcommand{\redtwo}[1]{\textcolor{black}{#1}}
\usepackage[caption=false]{subfig}
\begin{document}
\title{CHIME/FRB/Pulsar discovery of a nearby long period radio transient with a timing glitch}
\author[0000-0003-4098-5222]{Fengqiu Adam Dong}
  \affiliation{National Radio Astronomy Observatory, 520 Edgemont Rd, Charlottesville, VA 22903, USA}
  \affiliation{Department of Physics and Astronomy, University of British Columbia, 6224 Agricultural Road, Vancouver, BC V6T 1Z1 Canada}
\author[0000-0001-6812-7938]{Tracy E Clarke}
  \affiliation{U.S. Naval Research Laboratory, 4555 Overlook Ave., SW Washington, DC 20375}
\author[0000-0002-8376-1563]{Alice Curtin}
  \affiliation{Department of Physics, McGill University, 3600 rue University, Montr\'eal, QC H3A 2T8, Canada}
  \affiliation{Trottier Space Institute, McGill University, 3550 rue University, Montr\'eal, QC H3A 2A7, Canada}
\author[0009-0002-0330-9188]{Ajay Kumar}
  \affiliation{National Centre for Radio Astrophysics, Post Bag 3, Ganeshkhind, Pune, 411007, India}
\author[0000-0001-7348-6900]{Ryan Mckinven}
  \affiliation{Department of Physics, McGill University, 3600 rue University, Montr\'eal, QC H3A 2T8, Canada}
  \affiliation{Trottier Space Institute, McGill University, 3550 rue University, Montr\'eal, QC H3A 2A7, Canada}
\author[0000-0002-6823-2073]{Kaitlyn Shin}
  \affiliation{MIT Kavli Institute for Astrophysics and Space Research, Massachusetts Institute of Technology, 77 Massachusetts Ave, Cambridge, MA 02139, USA}
  \affiliation{Department of Physics, Massachusetts Institute of Technology, 77 Massachusetts Ave, Cambridge, MA 02139, USA}
\author[0000-0001-9784-8670]{Ingrid Stairs}
  \affiliation{Department of Physics and Astronomy, University of British Columbia, 6224 Agricultural Road, Vancouver, BC V6T 1Z1 Canada}
\author{Charanjot Brar}
  \affiliation{Department of Physics, McGill University, 3600 rue University, Montr\'eal, QC H3A 2T8, Canada}
  \affiliation{Trottier Space Institute, McGill University, 3550 rue University, Montr\'eal, QC H3A 2A7, Canada}
\author[0000-0002-7226-836X]{Kevin Burdge}
  \affiliation{Department of Physics, Massachusetts Institute of Technology, 77 Massachusetts Ave, Cambridge, MA 02139, USA}
\author[0000-0002-2878-1502]{Shami Chatterjee}
  \affiliation{Cornell Center for Astrophysics and Planetary Science, Cornell University, Ithaca, NY 14853, USA}
\author[0000-0001-6422-8125]{Amanda M.~Cook}
  \affiliation{Department of Physics, McGill University, 3600 rue University, Montr\'eal, QC H3A 2T8, Canada}
  \affiliation{Trottier Space Institute, McGill University, 3550 rue University, Montr\'eal, QC H3A 2A7, Canada}
\author[0000-0001-8384-5049]{Emmanuel Fonseca}
  \affiliation{Department of Physics and Astronomy, West Virginia University, PO Box 6315, Morgantown, WV 26506, USA }
  \affiliation{Center for Gravitational Waves and Cosmology, West Virginia University, Chestnut Ridge Research Building, Morgantown, WV 26505, USA}
\author[0000-0002-3382-9558]{B.~M.~Gaensler}
  \affiliation{Department of Astronomy and Astrophysics, University of California, Santa Cruz, 1156 High Street, Santa Cruz, CA 95060, USA}
  \affiliation{Dunlap Institute for Astronomy and Astrophysics, 50 St. George Street, University of Toronto, ON M5S 3H4, Canada}
  \affiliation{David A. Dunlap Department of Astronomy and Astrophysics, 50 St. George Street, University of Toronto, ON M5S 3H4, Canada}
\author[0000-0003-2317-1446]{Jason W. Hessels}
  \affiliation{Department of Physics, McGill University, 3600 rue University, Montr\'eal, QC H3A 2T8, Canada}
  \affiliation{Trottier Space Institute, McGill University, 3550 rue University, Montr\'eal, QC H3A 2A7, Canada}
  \affiliation{Anton Pannekoek Institute for Astronomy, University of Amsterdam, Science Park 904, 1098 XH Amsterdam, The Netherlands}
  \affiliation{ASTRON, Netherlands Institute for Radio Astronomy, Oude Hoogeveensedijk 4, 7991 PD Dwingeloo, The Netherlands}
\author[0000-0001-9345-0307]{Victoria M.~Kaspi}
  \affiliation{Department of Physics, McGill University, 3600 rue University, Montr\'eal, QC H3A 2T8, Canada}
  \affiliation{Trottier Space Institute, McGill University, 3550 rue University, Montr\'eal, QC H3A 2A7, Canada}
\author[0000-0002-5857-4264]{Mattias Lazda}
  \affiliation{Dunlap Institute for Astronomy and Astrophysics, 50 St. George Street, University of Toronto, ON M5S 3H4, Canada}
  \affiliation{David A. Dunlap Department of Astronomy and Astrophysics, 50 St. George Street, University of Toronto, ON M5S 3H4, Canada}
\author[0000-0002-7164-9507]{Robert Main}
  \affiliation{Department of Physics, McGill University, 3600 rue University, Montr\'eal, QC H3A 2T8, Canada}
  \affiliation{Trottier Space Institute, McGill University, 3550 rue University, Montr\'eal, QC H3A 2A7, Canada}
\author[0000-0002-4279-6946]{Kiyoshi W.~Masui}
  \affiliation{MIT Kavli Institute for Astrophysics and Space Research, Massachusetts Institute of Technology, 77 Massachusetts Ave, Cambridge, MA 02139, USA}
  \affiliation{Department of Physics, Massachusetts Institute of Technology, 77 Massachusetts Ave, Cambridge, MA 02139, USA}
\author[0000-0002-2885-8485]{James W.~McKee}
  \affiliation{Department of Physics and Astronomy, Union College, Schenectady, NY 12308, USA}
\author[0000-0001-8845-1225]{Bradley W.~Meyers}
  \affiliation{International Centre for Radio Astronomy Research (ICRAR), Curtin University, Bentley WA 6102, Australia}
  \affiliation{Australian SKA Regional Centre (AusSRC), Curtin University, Bentley WA 6102, Australia}
\author[0000-0002-8912-0732]{Aaron B.~Pearlman}
  \affiliation{Department of Physics, McGill University, 3600 rue University, Montr\'eal, QC H3A 2T8, Canada}
  \affiliation{Trottier Space Institute, McGill University, 3550 rue University, Montr\'eal, QC H3A 2A7, Canada}
  \affiliation{Banting Fellow}
  \affiliation{McGill Space Institute Fellow}
  \affiliation{FRQNT Postdoctoral Fellow}
\author[0000-0001-5799-9714]{Scott M.~Ransom}
  \affiliation{National Radio Astronomy Observatory, 520 Edgemont Rd, Charlottesville, VA 22903, USA}
\author[0000-0002-7374-7119]{Paul Scholz}
  \affiliation{Department of Physics and Astronomy, York University, 4700 Keele Street, Toronto, ON MJ3 1P3, Canada}
  \affiliation{Dunlap Institute for Astronomy and Astrophysics, 50 St. George Street, University of Toronto, ON M5S 3H4, Canada}
\author[0000-0002-2088-3125]{Kendrick M.~Smith}
  \affiliation{Perimeter Institute of Theoretical Physics, 31 Caroline Street North, Waterloo, ON N2L 2Y5, Canada}
\author[0000-0001-7509-0117]{Chia Min Tan}
  \affiliation{International Centre for Radio Astronomy Research (ICRAR), Curtin University, Bentley WA 6102, Australia}
  \affiliation{Australian SKA Regional Centre (AusSRC), Curtin University, Bentley WA 6102, Australia}
\newcommand{\allacks}{
F.A.D was support by the U.B.C Four Year Fellowship, and is now an NRAO Jansky Fellow.
Basic research in radio astronomy at the U.S. Naval Research Laboratory is supported by 6.1 Base funding. Construction and installation of VLITE was supported by the NRL Sustainment Restoration and Maintenance fund.
A.P.C. is a Vanier Canada Graduate Scholar. 
Pulsar and FRB research at UBC are supported by an NSERC Discovery Grant and by the Canadian Institute for Advanced Research. 
E.F., I.S., S.C., S.M.R. are members of the NANOGrav Physics Frontiers Center, supported by the NSF award 2020265.
A.M.C. is funded by an NSERC Doctoral Postgraduate Scholarship.
E.F. is supported by the NSF under grant number AST-2407399.
The Dunlap Institute is funded through an endowment established by the David Dunlap family and the University of Toronto. B.M.G. acknowledges the support of the Natural Sciences and Engineering Research Council of Canada (NSERC) through grant RGPIN-2022-03163, and of the Canada Research Chairs program.
V.M.K. holds the Lorne Trottier Chair in Astrophysics \& Cosmology, a Distinguished James McGill Professorship, and receives support from an NSERC Discovery grant (RGPIN 228738-13), from an R. Howard Webster Foundation Fellowship from CIFAR, and from the FRQNT CRAQ.
K.W.M. holds the Adam J. Burgasser Chair in Astrophysics
A.B.P. is a Banting Fellow, a McGill Space Institute~(MSI) Fellow, and a Fonds de Recherche du Quebec -- Nature et Technologies~(FRQNT) postdoctoral fellow.
S.M.R. is a CIFAR Fellow. 
P.S. acknowledges the support of an NSERC Discovery Grant (RGPIN-2024-06266).
K.S. is supported by the NSF Graduate Research Fellowship Program.
}

\begin{abstract}
  We present the discovery of a 421\,s long period radio transient (LPT) using the CHIME telescope, CHIME J0630+25. The source is localized to RA=06:30:38.4$\pm1'$ Dec=25:26:24$\pm1'$ using voltage data acquired with the CHIME baseband system. A timing analysis shows that a model including a glitch is preferred over a non-glitch model with $dF/F=1.3\times10^{-6}$, consistent with other glitching neutron stars. The timing model suggests a surface magnetic field of $\sim1.5\times10^{15}$\,G and a characteristic age of $\sim1.28\times10^{6}$\,yrs. A separate line of evidence to support a strong local magnetic field is an abnormally high rotation measure of $RM=-347.8(6)\, \mathrm{rad\, m^{-2}}$ relative to CHIME J0630+25's modest dispersion measure of 22(1)\,pc cm$^{-2}$, implying a dense local magneto-ionic structure. As a result, we believe that CHIME J0630+25 is a magnetized, slowly spinning, isolated neutron star. This marks CHIME J0630+25 as the longest period neutron star and the second long period neutron star with an inferred magnetar-like field. Based on dispersion measure models and comparison with pulsars with distance measurements, CHIME J0630+25 is located at a nearby distance of 170$^{+310}_{-100}$~pc (95.4\%), making it an ideal candidate for follow-up studies.
\end{abstract}


  \section{Introduction}
\label{sec:intro} Recently, a new class of radio-emitting objects known as long-period transients (LPTs)\footnote{Note that these sources have also been given some other names in the literature, such as ultra long period transients (ULPT) or long period radio transients (LPRT). In this study, we use the term long period transient throughout.} has been discovered. LPTs are characterised by their exceptionally long periods, wide burst widths (up to 60 seconds), and complex temporal and spectral microstructure. Two types of sources, white dwarfs and neutron stars, have emerged as the favoured models for LPTs due to their emission characteristics and rotation periods. 

Radio pulsars are the most common form of detectable neutron star. They are remarkably accurate celestial clocks. By carefully measuring pulse times of arrival (TOAs), we can fully account for every rotation of a pulsar through a process called pulsar timing. The complex temporal and spectral structure of LPT radio emission is similar to that of magnetars, a subset of high magnetic field neutron stars that share many similarities with radio pulsars. Some have argued that this favours a neutron-star model for LPTs \citep[e.g.][]{Beniamini:wadiasingh:hare:2023}. The longest period pulsars, PSR J0250+5854, and J0311+1402 have a rotational period of 23.5\,s, and 40.9~s respectively \citep{Tan:Bassa:Cooper:2018,Wang:uttarkar:Shannon:2025}. In addition, there is one 76\,s neutron star with magnetar-like field strength, \meerkat{} \citep{caleb:heywood:rajwade:2022}. Some consider \meerkat{} an LPT. Finally, there is one magnetar candidate, 1E 161348-5055, with a period of 6.67~h at the centre of supernova remnant RCW103 \citep{deluca:caraveo:mereghetti:2006}. However, only high-energy emission has been detected from 1E 161348-5055 \citep{Gotthelf:Petre:Hwang:1997,Dai:Evans:burrows:2016}.

LPTs range from minute \citep{caleb:heywood:rajwade:2022} to hour timescales \citep[e.g][]{Hurley-Walker:Rea:2023,deRuiter:2024,Wang:Rea:Bao:2024,Li:Yuan:Wu:2024}. Some have been identified to be white dwarf (WD) M-Dwarf binaries via optical spectroscopy, like \ILT{} \citep{deRuiter:2024} and \gleamtwo{} \citep{Hurley-Walker:2024,Rodriguez:2025}. On the other hand, as previously mentioned, \meerkat{} is a confirmed neutron star LPT. This is supported by the low timing residuals (r.m.s. residual of 7.5$\times10^{-5}$ in phase units) and its apparent isolated nature when pulsar timing techniques are used. Isolated WDs have never been observed to emit coherent pulsed radio emission. Furthermore, the quasiperiodic substructure of the \meerkat{} bursts directly aligns it with other neutron stars in a universal quasiperiod-period relationship for neutron stars proposed by \cite{kramer:liu:2023}.

This study details the discovery of \J{}, a nearby 421\,s LPT. Section \ref{sec:observations} details the observations made. Section \ref{sec:timing} details the timing methodology and results. Section \ref{sec:polarisation} details the polarization analysis. Finally, Section \ref{sec:Discussion} provides a discussion of the main results of this study.

  \section{Observations}
\label{sec:observations} 
CHIME/FRB is a trigger-based FRB-detection instrument on the CHIME telescope \citep{10.3847/1538-4365/ac6fd9} that constantly scans the overhead sky with 1024 FFT-formed beams between 400--800~MHz and a field of view of approximately $2^{\circ}$ in RA and $100^{\circ}$ in Dec \citep{10.3847/1538-4357/aad188}. The instrument triggers on impulsive signals \citep{10.3847/1538-4357/aad188}. Then, machine learning software will determine if the incoming signal is terrestrial or astrophysical. Data are then saved to disk if a certain S/N threshold (currently 8.5) is met. Due to the substantial data volume, data for sources within the Milky Way Galaxy were not saved by CHIME/FRB until October 2022. However, metadata was saved for each ``event'' regardless of origin. The metadata contains real-time pipeline-derived information such as the right ascension (RA), Declination (Dec), dispersion measure (DM), TOA, and signal-to-noise (S/N). We used the CHIME Metadata Clustering Analysis (CHIMEMCA), to identify CHIME J0630+25 \citep[see][for a full description]{dong:crowter:meyers:2023}. A cluster was identified with the first burst on MJD 58772. On MJD 60463, we made a detection with the CHIME/FRB system with channelized raw voltage (baseband) data. This allows improved localization and polarization analysis \citep{2021ApJ...910..147M, 10.3847/1538-4357/ad464b}. The best fit localization of CHIME J0630+25 with baseband data is RA=06:30:38.4$\pm1'$ Dec=25:26:23$\pm1'$. By comparing the DMs to other pulsars in the ATNF pulsar catalog and the YMW16 electron density model \citep{10.3847/1538-4357/835/1/29}, we determined the distance to \J{} to be 170$^{+310}_{-100}$~pc. YMW16 model was used instead of NE2001 \citep{10.48550/arXiv.astro-ph/0207156} as it has been shown that YMW16 is better for nearby pulsars \citep{10.1017/pasa.2021.33}. Details regarding the distance determination procedure are given in Appendix \ref{dmdist}

\begin{figure}[h]
  \centering
  \includegraphics[width=\textwidth]{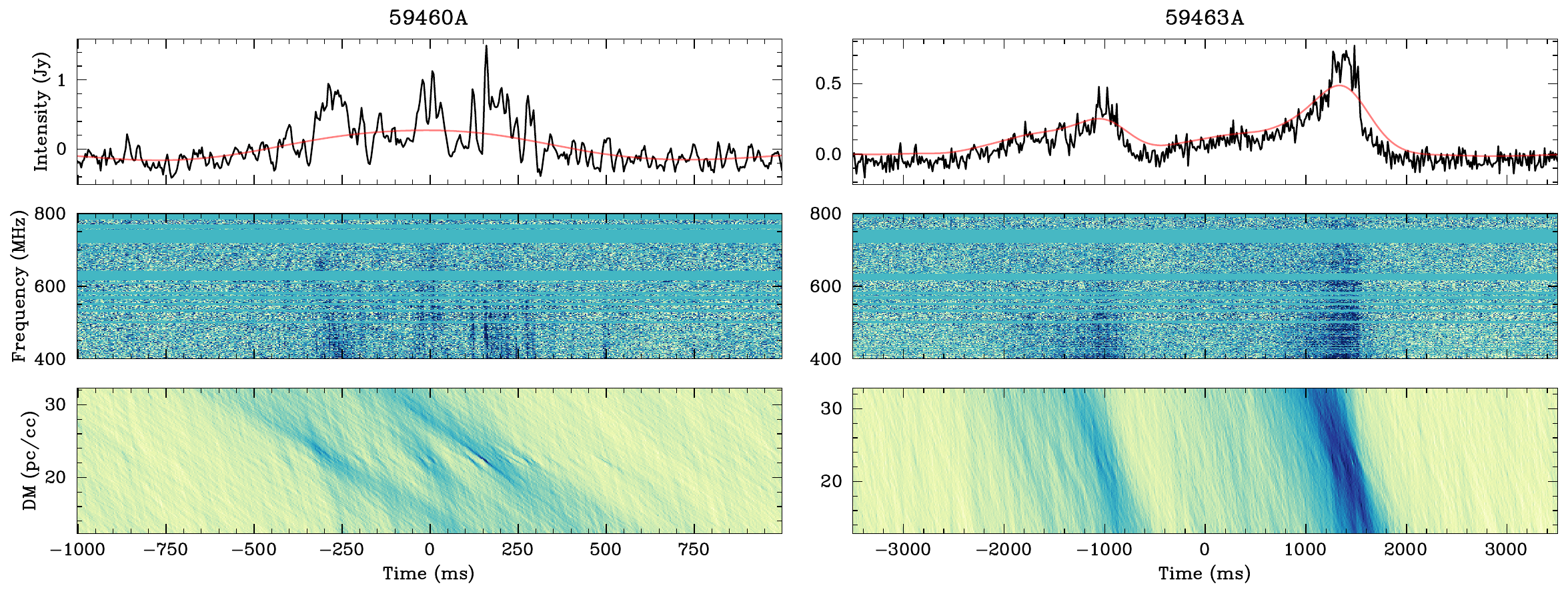}
  \caption{Two pulses from CHIME J0630+25 detected by CHIME/Pulsar. The top panel for each burst contains frequency averaged and dedispersed time series. \red{The red line is the smoothed profile. The plots are not necessarily centered on the extracted TOA; rather, the center is chosen to best show the burst morphology.} The second panel shows the dynamic spectrum of each burst, and the bottom panel shows the dedispersion heat map for each burst. The dedispersion heat map shows the power for many different DM trials. All astrophysical pulses of CHIME J0630+25 pulses centre on DM$\approx 22~$pc~cm$^{-3}$.}
  \label{fig:bursts_1}
\end{figure}

The CHIME/Pulsar instrument forms ten steerable phased array tracking beams to track sources as they pass through the CHIME field of view. It produces high-time resolution spectra, packaged as conventional SigProc-style filterbank data\footnote{\url{https://sigproc.sourceforge.net/}}. CHIME/Pulsar can also correct for intrachannel dispersion smearing through coherent dedispersion and can record data at significantly higher time resolutions than CHIME/FRB. These advantages contribute to increased sensitivity for CHIME/Pulsar compared to CHIME/FRB. Follow-up observations were conducted nearly daily, consisting of about $\sim$10 minutes per scan. These observations remain ongoing. With CHIME/FRB, we detected eight bursts above S/N 8.5, where one burst had baseband voltage data saved and the rest had only metadata. With CHIME/Pulsar, we detected 15 bursts with Stokes-I intensity data showing extended widths and complex structures. Due to two same-day detections from CHIME/Pulsar we determined an initial period of $\sim421$\,s. All bursts are detailed in Table \ref{tab:burst_characteristics_1}. \red{The dynamic spectra for the CHIME/Pulsar detections are shown in Figure \ref{fig:bursts_1} and Appendix \ref{sec:followup}.} We confirmed the CHIME/FRB bursts with only metadata to be real astrophysical events as the arrival time is consistent with the rotation period of \J{}.

We flux calibrate the CHIME/Pulsar data using observations of 3C133. These calibration observations serve to determine the system equivalent flux density (SEFD) of the telescope at the declination of \J{}. As CHIME/Pulsar observations can have a varying baseline on timescales of seconds, we attempt to subtract off the baseline with a fit to the non-bursting segment of the data. We then integrate the burst over its duration to obtain the fluence, $\textrm{F}$. The effective width, $W_{\text{eff}}$, is defined as the fluence divided by the peak flux density. The spectral index is calculated by fitting a power law of the form $S(\nu) = A\nu^{\alpha}$ to the spectrum of the burst, where $S(\nu)$ is the flux density, $\nu$ is the frequency, and $A$ is the amplitude parameter. A detailed discussion of the calibration routine is given in the Appendix \ref{calibration}. 

In addition, we observed the source location with the Green Bank Telescope (GBT) for 16 hours and the upgraded Giant Metrewave Radio Telescope (uGMRT) for 12 hours. Unfortunately, the uGMRT data was unusable due to radio frequency interference.  Furthermore, we also searched archival Very Large Array Low-band Ionosphere and Transient Experiment (VLITE) data. No additional bursts were detected. We also monitored the source location of \J{} using the Niels Geherels {\it Swift} Telescope for 32\,ks. No sources were detected within the $3\sigma$ baseband localization region of \J{}. We place a $3\sigma$ unabsorbed flux upper limit of 1.2$\times$10$^{-14}$ erg cm$^{-2}$ s$^{-1}$ and 3.2$\times$10$^{-14}$ erg cm$^{-2}$ s$^{-1}$ for a blackbody and power-law spectrum, respectively. \red{Finally, an optical counterpart was searched for with periods around 421\,s. Unfortunately, there were $\sim150$ marginal candidates. Visual inspection of the light curves reveals that the data are noisy and likely contain mostly false positives.} Details regarding these observations are provided in the Appendix \ref{observations}. 

\begin{table}[ht] 
  \centering
  \caption{Detection details for each burst. The TOAs are given in the topocentric reference frame at CHIME.}
  \centering
  \begin{tabular}{cllcccc}
    \hline
    \hline
    Burst        & Instrument   & TOA              & $W_{\text{eff}}$ & $\textrm{F}$      & $\alpha$  & DM           \\
                 &              & MJD              & ms               & Jy~ms    &           & pc~cm$^{-3}$ \\
    \hline
    \hline
    58772A$^{\&}$       & CHIME/FRB    & 58772.538085(12) & --               & --       & --        & 23(3)        \\
    58855A$^{\&}$       & CHIME/FRB    & 58855.307944(12) & --               & --       & --        & 23(3)        \\
    58860A$^{\&}$       & CHIME/FRB    & 58860.296981(12) & --               & --       & --        & 24(3)        \\
    58871A$^{\&}$       & CHIME/FRB    & 58871.265266(12) & --               & --       & --        & 24(3)        \\
    59167A$^{\&}$       & CHIME/FRB    & 59167.453835(12) & --               & --       & --        & 24(3)        \\
    59341A       & CHIME/Pulsar & 59341.973144(12) & 600(230)         & 800(240) & -1.2(3)   & 20(3)        \\
    59341B       & CHIME/Pulsar & 59341.978026(12) & 150(90)          & 60(20)   & -0.05(30) & 23(3)        \\
    59456A       & CHIME/Pulsar & 59456.659903(12) & 100(100)         & 60(20)   & -1.4(3)   & 23(3)        \\
    59456B       & CHIME/Pulsar & 59456.664743(12) & 180(120)         & 90(30)   & -2.0(3)   & 23(3)        \\
    59460A$^*$       & CHIME/Pulsar & 59460.648754(12) & 210(90)          & 400(120) & -2.7(3)   & 20(3)        \\
    59463A       & CHIME/Pulsar & 59463.642824(36) & 950(270)         & 680(200) & -2.5(3)   & 22(6)        \\
    59548A       & CHIME/Pulsar & 59548.413841(12) & 370(140)         & 190(60)  & -3.9(3)   & 20(3)        \\
    59553A$^{*}$ & CHIME/Pulsar & 59553.397700(12) & 330(140)         & 430(130) & -3.4(3)   & 20(7)        \\
    59563A       & CHIME/Pulsar & 59563.370419(12) & 600(200)         & 500(150) & -2.6(3)   & 23(3)        \\
    59565A       & CHIME/Pulsar & 59565.364998(12) & 250(60)          & 130(40)  & -2.9(3)   & 21(3)        \\
    59574A       & CHIME/Pulsar & 59574.343036(12) & 650(60)          & 370(100) & -2.5(3)   & 25(4)        \\
    60173A       & CHIME/Pulsar & 60173.696625(12) & 30(10)           & 40(10)   & -2.9(3)   & 25(4)        \\
    60372A       & CHIME/Pulsar & 60372.156479(12) & 23(10)           & 12(8)    & -5.2(3)   & 25(4)        \\
    60376A       & CHIME/Pulsar & 60376.146068(12) & 230(60)          & 300(100) & -1.5(3)   & 25(4)        \\
    60463A$^{*}${}\textsuperscript{\textdagger}& CHIME/Pulsar & 60463.906535(12) & 190(200)         & 240(70) & -5.4(3)   & 23(3)        \\
    \hline
  \end{tabular}
  
\label{tab:burst_characteristics_1}
  \red{$^{\&}$ These detections are metadata only.}\\
  $^{*}$ These pulses were simultaneously detected with both the CHIME/Pulsar and CHIME/FRB systems.\\
  \red{\textsuperscript{\textdagger} This burst had raw voltage baseband data saved.}\\
\end{table}

  \section{Timing}\label{sec:timing}
\begin{figure}[ht]
  \centering
  \gridline{\fig{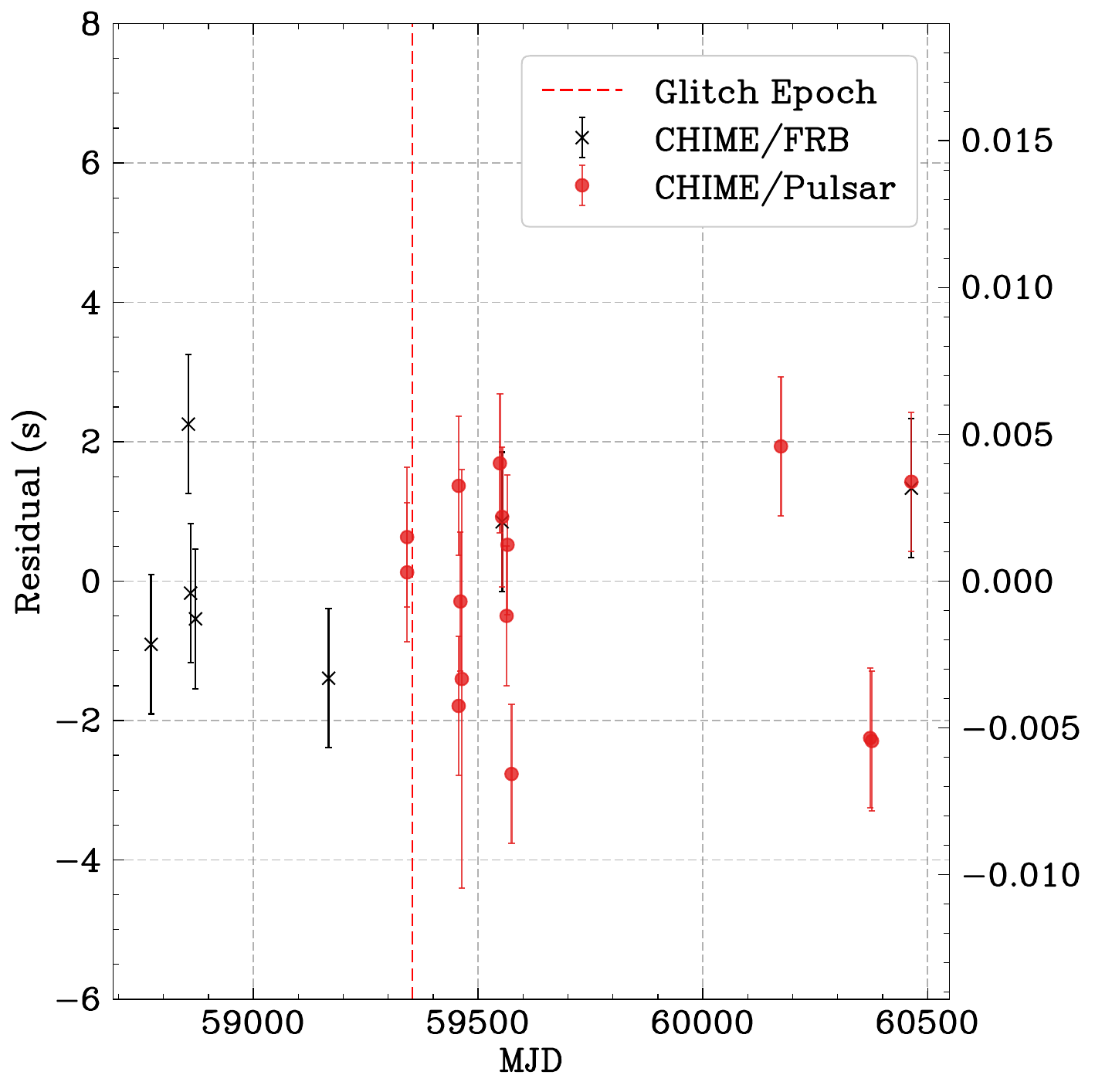}{0.49\textwidth}{(a)} \fig{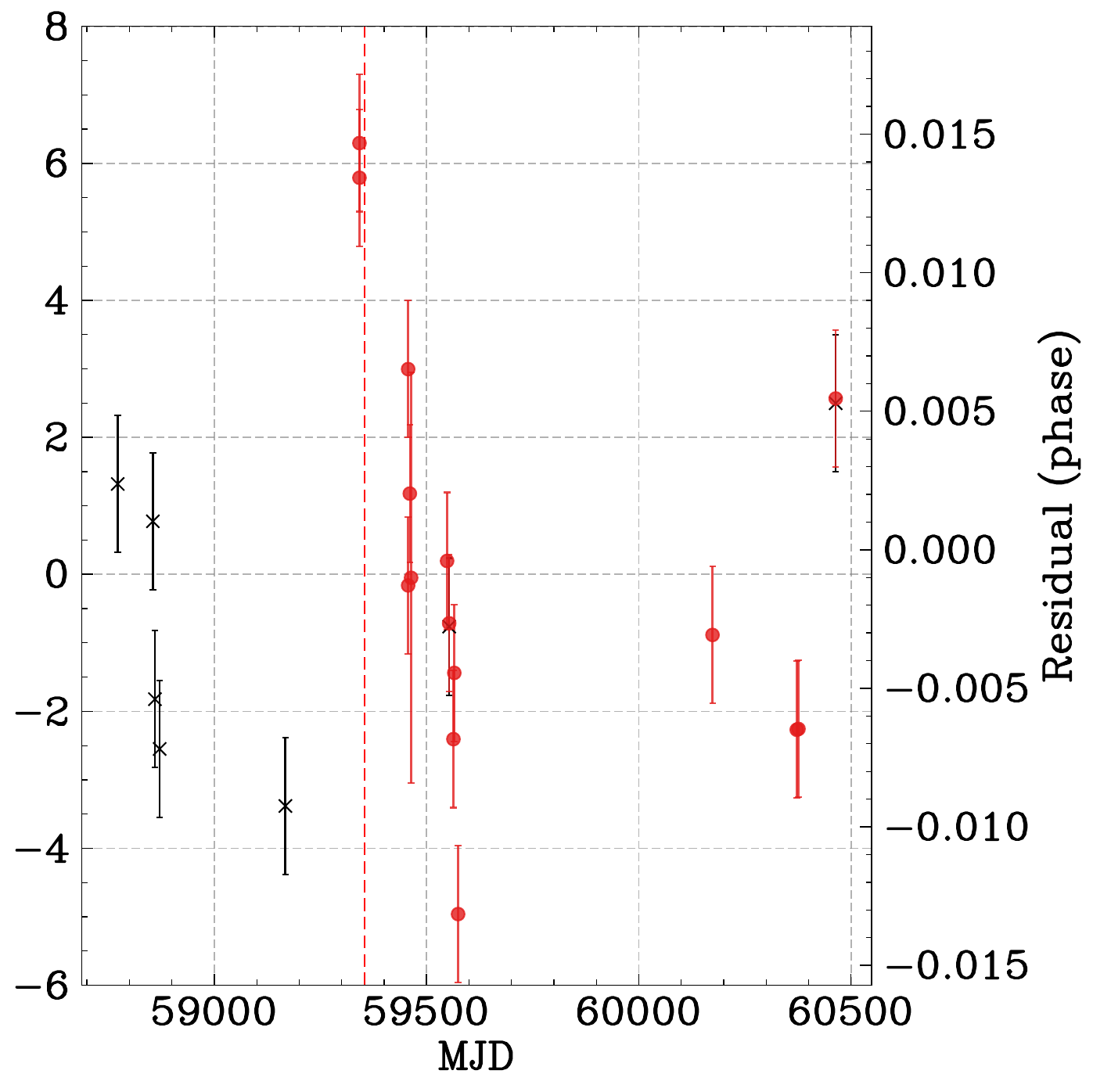}{0.49\textwidth}{(b)}}
  \gridline{\fig{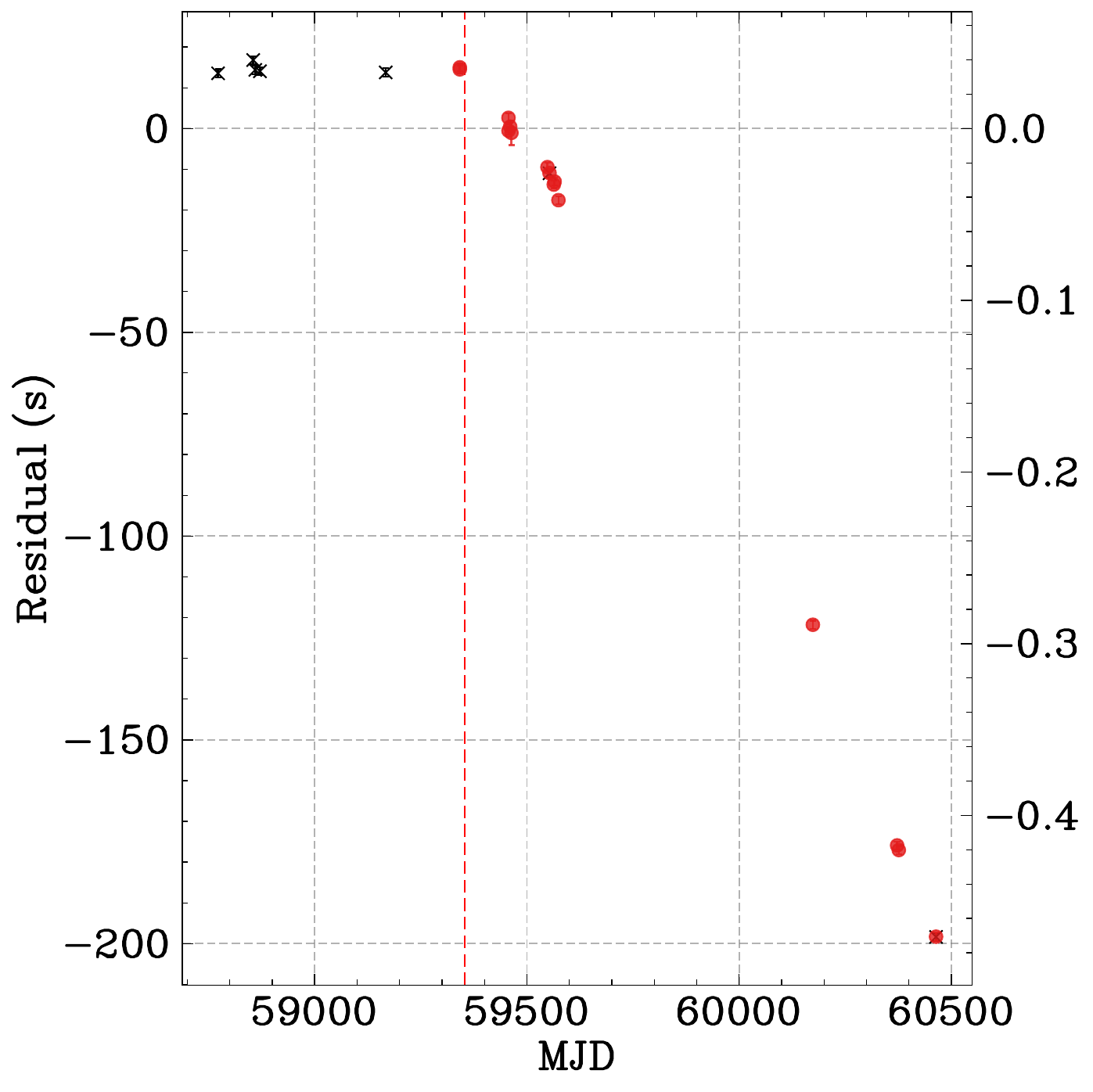}{0.49\textwidth}{(c)} \fig{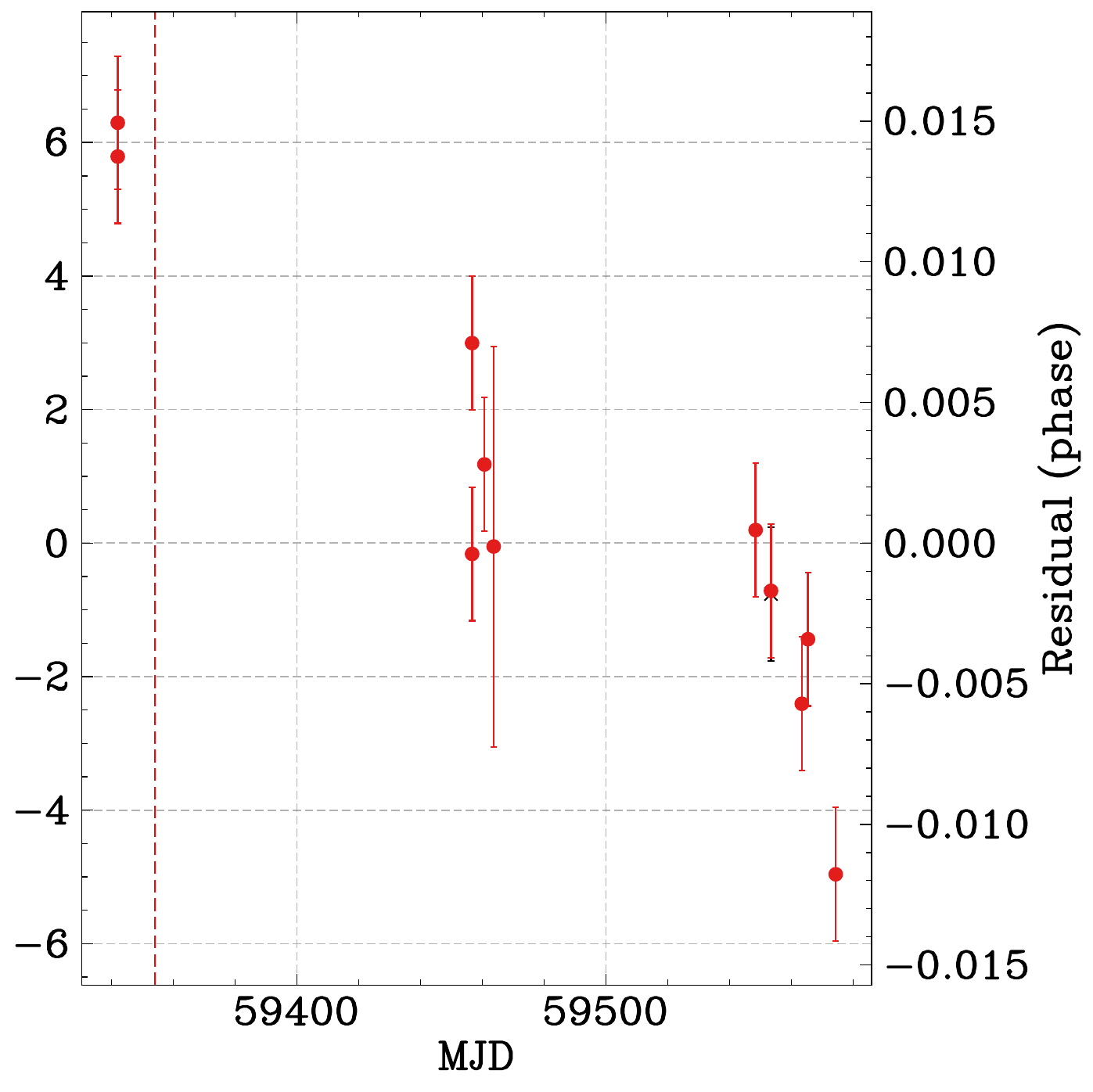}{0.49\textwidth}{(d)}}
  \caption{\redtwo{No EFACs were included in any of the timing residual plots. }(a) Timing residuals using the glitch model. (b) Timing residuals using
  the non-glitch model. (c) Timing residuals obtained by fitting only to the
  bursts before the glitch epoch. (d)
  Zoomed in timing residuals of panel (b) showing a clear downward sloping
  structure.}
  \label{fig:glitch}
\end{figure}

\begin{table}[ht]
  \caption{We do not fit for the Right Ascension or the Declination as they are
  fixed at the baseband localised coordinates. We do not fit F2 for the ``No Glitch''
  solution as it does not improve the fit. \redtwo{The errors provided are all given with an EFAC of 1.6, thereby reducing the $\chi_{reduced}=1$.}}
  \label{tab:J0630+25_timing}
  \centering
  \begin{tabular}{lllll}
    \hline
    \hline
    Property                                  & Glitch             & No Glitch          & Post Glitch Fit $^*$\\
    \hline
    \hline
    R.A. (hh:mm:ss)                           & 06h30m38.40s       & 06h30m38.40s       & 06h30m38.40s       \\
    Dec (dd:mm:ss)                            & 25$^{\circ}$26'23" & 25$^{\circ}$26'23" & 25$^{\circ}$26'23" \\
    $P$(s)                                    & 421.35584(8)       & 421.355420(6)      & 421.3551(2)        \\
    $\dot{P}(\times10^{-13}~\text{ss}^{-1})$  & 52(11)              & --7.8(1.4)           & 80(64)             \\
    $\ddot{P}(\times10^{-19}~\text{s}^{-1})$  & --1.1(5)           & --                 & --0.9(8)           \\
   Glitch Epoch (GLEP)                       & 59354(14)       & --                 & --                 \\
    Glitch Frequency (GLF0) $10^{-9}$~Hz      & 3.08(56)           & --                 & --                 \\
    NTOA                                      & 23                 & 23                 & 16                 \\
    JUMP                                      & 0.4(4)            & 0.4(4)            & 0.4(4)            \\
    RMS Residuals (s)                         & 1.42               & 2.73               & 1.6                \\
    RMS Residuals (phase)                     & 0.0034             & 0.0065             & 0.0038             \\
    $\chi_{reduced}^2$$^{\dagger}$                          & 2.6                & 8.2                & 3.5                \\
    \hline
    Derived Values                            &                    &                    & \\
    \hline
    Galactic Longitude (deg)                  & 187.92             & 187.92             & 187.92             \\
    Galactic Latitude (deg)                   & 7.07               & 7.07               & 7.07               \\
    $\tau$ ($\times 10^{6}$~yr)               & 1.28               & --                 & 0.8                \\
    B$_{\text{surface}}$ ($\times 10^{15}$~G) & 1.5                & --                 & 1.8                \\
    $\dot{E}$ ($\times 10^{27}$~erg/s)        & 2.76               & --                 & 4.1                \\
    \hline
  \end{tabular}\\
  $^{*}$ Here we fit only data after the glitch epoch. All of these TOAs are
  derived from total intensity.\\
  \redtwo{$^{\dagger}$ Here, we are providing the raw $\chi_{reduced}^2$ without any EFAC for comparison purposes. For all errors in the timing solution, we set EFAC=1.6, commensurate with a $\chi_{reduced}^2=1$ for the glitch solution.}
\end{table}

Time of arrival (TOA) extraction is vital for period determination and pulsar
timing. This is necessary for both the CHIME/FRB and CHIME/Pulsar datasets.
Unfortunately, for most of the CHIME/FRB detections, the only data products saved were
the metadata. These contain a timestamp for the pulse peak but no total intensity
data, and therefore, it is difficult to know if the system triggered on one of the
microstructure peaks. Therefore, we took a conservative TOA error of 1~s for all
metadata-only detections. The CHIME/FRB TOAs are referenced at the bottom of the
CHIME band at 400~MHz. For the CHIME/Pulsar detections where the total intensity
data are recorded, we followed the prescription outlined in
\cite{Hurley-Walker:Rea:2023}. Specifically, we smoothed the bursts' pulse profile
using a variable Gaussian kernel. The maximum of the smoothed profile is taken as the
TOA, and the full-width-half-max (FWHM) is taken as the uncertainty of the TOA. \red{Furthermore, we set a lower limit of 1\,s on the errors due to the variability of the pulse shapes.} \red{We note that there can be other ways to determine the TOA, for example, one could consider using the effective width or an S/N weighted TOA. However, as the pulse-to-pulse variability is large, we believe that the smoothed FWHM provides a conservative estimate of the TOA. As shown in Table \ref{tab:J0630+25_timing}, the reduced $\chi^{2}$ suggests that the jitter is only slightly larger than the pulse width (about a factor of 1.61), which is expected for a single-pulse timing analysis. We find no obvious systematic or secular changes in the pulse shape.}
The CHIME/Pulsar TOAs are referenced at the top of the CHIME band at 800~MHz.

Our initial period guess was $\sim 421$\,s due to the two days where we detected
two consecutive bursts, 59341A, B, and 59456A, B. We then used \texttt{Astropy}
to correct the TOAs to the reference frame of the solar system barycentre and
used the \texttt{rrat\_period} module of \texttt{PRESTO} to confirm that the
most likely period for \J{} is indeed $\sim421$~s.

We created a timing ephemeris by taking the 421\,s periodicity as the starting
point. \red{Compared to TOAs of a pulsar, the data is sparsely sampled due to the sporadic nature of \J{}. However, we note that the period of \J{} is $\sim$400 times larger than standard ``slow’’ pulsars and even greater for millisecond pulsars. Therefore, phase-coherent timing can be much more forgiving of data gaps, as there will have been many fewer rotations within the gaps when compared to pulsars. For perspective, a data gap of 1 year for \J{} is equivalent to a data gap of 0.87 days for a 1\,s pulsar.} We included a fixed 0.452\,s JUMP during the timing fit between the CHIME/FRB
and CHIME/Pulsar TOAs. This was determined by the two bursts simultaneously detected
by both instruments, as shown in Table \ref{tab:burst_characteristics_1}. We tested
the validity of the phase jump between the two instruments by performing the same
TOA extraction procedure described above on a slow pulsar, namely PSR J0012+54, and found the jump to be largely consistent. We provide the timing residuals of J0012+54 in Appendix \ref{sec:appen_J0012}. We then used \texttt{PINT}\footnote{https://github.com/nanograv/PINT} \citep{luo:ransom:demorest:2021} and \texttt{TEMPO2} \footnote{https://ascl.net/1210.015} \citep{Edwards:hobbs:Manchester:2006},
to perform a least squares fit to all the TOAs of \J{}, holding the position fixed
at the baseband localized value. \red{To test the robustness of this solution, we perform a large-scale grid search over the frequency and frequency derivative. We find no other minimums other than the one identified by the least squares fitting routine. This is shown in Appendix \ref{sec:appen_timing}.}

Our preferred solution incorporates a glitch -- an abrupt increase in the pulsar
spin frequency -- near epoch MJD 59354. \red{In the glitch fit, we left the glitch epoch and glitch frequency as free parameters with initial conditions near the beginning of the secular downward drift beginning at burst 59341 A, B. If we were to fit only around or after the glitch epoch, we obtain a timing solution consistent with only the glitch model, with similar $\chi_{reduced}$. That is, a timing solution with a spinning-down $\dot{P}$, as expected for an isolated pulsar. This rules out a situation like that of small orbit white dwarfs \citep{Screiber:Belloni:Gansicke:2021}, where the central engine is accreting angular momentum from a companion. Of course, due to the reduction in data and timing baseline, the significance of the detection is reduced. We also present this timing solution in Table \ref{tab:J0630+25_timing}.}
The fractional glitch amplitude $|dF/F|=1.3\times10^{-6}$, consistent with the
range observed in other pulsars from $10^{-9}-10^{-3}$ \citep{McKee:Janssen:Strappers:2016,Serim:Sahiner:Cerri-Serim:2017}.
The results for the fit are provided in Table \ref{tab:J0630+25_timing}, and the
residuals are provided in Figure \ref{fig:glitch}. Figure \ref{fig:glitch} shows,
for comparison, fits to the entire data set without a glitch (b), and a fit to only
the pre-MJD-59354 data (c). The non-glitch model has much larger residual
scatter, with r.m.s. values of 1.45\,s vs 2.74\,s for the glitch vs non-glitch
models respectively. The non-glitch timing solution includes a negative
$\dot{P}$. \redtwo{However, we note that a negative $\dot{P}$ is only unphysical for isolated objects. Clear counter examples are cataclysmic binaries \citep[e.g.][]{paice:scaringi:segura:2024} and the recent discovery of CHIME/ILT J1634+44 \citep{Bloot:2025,Dong:2025}}. Furthermore, the non-glitch timing solution has significant residual structure. In contrast, the
timing solution with a glitch has a positive $\dot{P}$ and random residuals. We show
the structure in panel (d) of Figure \ref{fig:glitch}. We further test the
significance of the structure with the Pearson correlation coefficient. The non-glitch
model has a correlation coefficient of -0.913 with a p-value of $8\times10^{-5}$,
i.e., the trend in Figure \ref{fig:glitch} (d) is significant. For the same data
with the glitch model, we obtain a correlation coefficient of -0.11 with a p-value
of 0.73, i.e., any trend in the data is insignificant. Therefore, the glitch model
has produced the desired outcome of random residuals, while the non-glitch model has not.

To test the significance of the glitch model, we performed an F-test for the
glitch epoch (GLEP), glitch frequency (GLF0), and the second frequency
derivative (F2) parameters. This resulted in a highly significant F-statistic value
of $4.9\times10^{-5}$. A glitch model with only GLEP and GLF0 and no second
frequency derivative resulted in a significant F-test statistic of
$7.6\times10^{-5}$ relative to the non-glitch model. This suggests that,
regardless of whether F2 is included, the glitch model is significantly preferred
over the non-glitch model. To test the significance of the F2 parameter, we performed
an F-test using the glitch model with and without the F2 parameter. This
resulted in a F-test statistic of $0.0069$. This suggests that the F2 parameter is
preferred for the glitch model. 

Finally, as previously discussed, we performed a fit to only the TOAs after
the glitch epoch. This resulted in a timing solution consistent with the
glitch model, with a positive $\dot{P}$, albeit at lower significance. We note that if we opt not to fit F2, the significance increases drastically and is only consistent with the glitch model; however, we opt to keep F2 for comparison purposes. All TOAs and timing solutions are
provided as a data resource.

In summary, a glitch model is preferred for four reasons:
\begin{enumerate}
  \item The non-glitch model has a factor of two larger r.m.s. residuals.

  \item The non-glitch model contains a significant downward sloping structure in the
    residuals, while the glitch model is randomly scattered.

  \item The F-test statistic is significant for the glitch epoch and glitch frequency
    parameters.

  \item \red{Fitting a standard non-glitch model to data only before or after the glitch results in a spinning down (positive $\dot{P}$), with a similar inferred field strength as the glitch model.}

\end{enumerate}
\red{Our best-fit timing solution marginally prefers a $\ddot{P} \ne 0$. In principle, this second-order variation can arise from several mechanisms, including stochastic variations due to ``timing noise" \citep{Antonelli:Basu:Haskell:2023} and/or orbital motion \citep{Joshi:Rasio:1997}. However, the marginal estimate of $\ddot{P}$ prevents a clear interpretation. Additional observations will meaningfully constrain the likely sources of rotational evolution beyond spin-down.}
Our timing solution results in a surface magnetic field of $B_{\text{surface}}
=1.5\times10^{15}$~G, and a characteristic age of $\tau=1.3\times10^{6}$~yr. This
is consistent with a long-lived magnetar model.

  \section{Polarization}
\label{sec:polarisation}

Only a single burst from \J{} had baseband raw voltage data saved. We analyzed the polarized signal using methods/routines commonly applied to studying FRBs detected through the CHIME/FRB backend \citep{Mckinven2023,Mckinven2023b}. First, we converted the raw, beamformed data into the four Stokes parameters $-$ $I$ (total intensity), $Q$ and $U$ (linear polarization components), and $V$ (circular polarization) $-$ using standard techniques \citep{Mckinven2021}. These polarization spectra were extracted by temporally integrating the signal over the brightest subcomponent of the signal.

\begin{figure}[ht]
  \centering
  \includegraphics[width=1.0\textwidth]{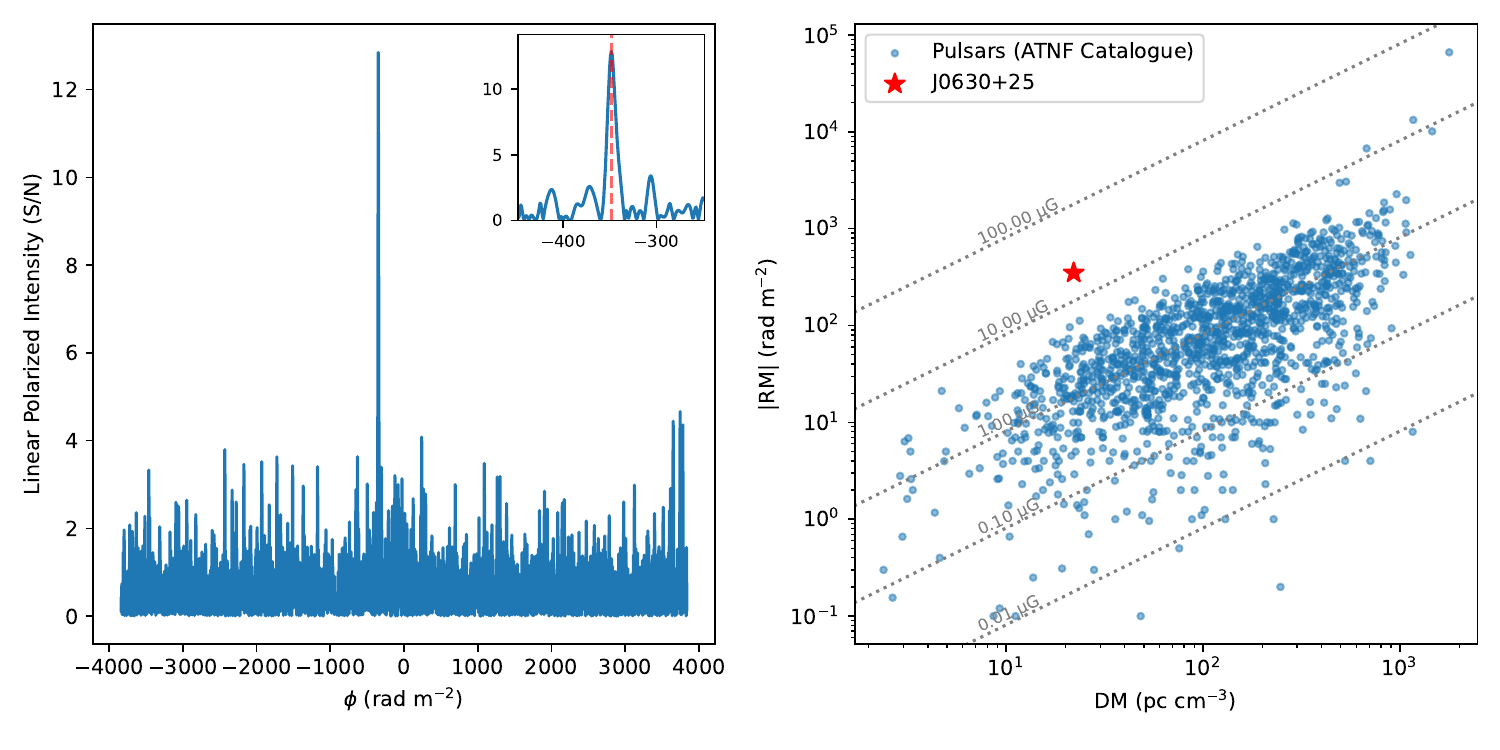}
  \caption{Left: The Faraday dispersion function (FDF) of the single polarized detection of \J{} showing linear polarized intensity as a function of trial Faraday depths, $\phi$. Inset plot highlights the FDF peak near an $\mathrm{RM}\sim -350\; \mathrm{rad\, m^{-2}}$. Right: DM--$\mathrm{|RM|}$ distribution of Galactic pulsar sample (blue points; ATNF Catalogue) versus equivalent measurement from \J{} (red star). Diagonal gray lines indicate different magnetic field strengths (0.01,0.1,1,10,100 $\mathrm{\mu G}$) determined from the relation, $B_{\parallel}=1.23\frac{|\mathrm{RM}|}{\mathrm{DM}} \; \mathrm{\mu G}$}.
  \label{fig:rm_vs_dm}
\end{figure}

To measure how the polarized signal was affected by magnetized plasma along its path, a phenomenon known as Faraday rotation, we applied a parametric $QU$-fitting technique. This technique enables astrophysical signals to be effectively parsed from contaminant signals from the instrument \citep{Mckinven2021}. An $RM=-347.8 \pm 0.6\; \mathrm{rad\, m^{-2}}$ was found to be consistent with the value obtained independently using the Faraday dispersion function (FDF), a complementary technique based on Fourier analysis of the polarized signal. \red{Figure~\ref{fig:rm_vs_dm} (left panel) summarizes the RM detection, displaying the FDF of the burst where a significant peak (S/N$\gtrsim$12) is visible at the nominal RM of the event. While instrumental effects are known to produce artefacts in the FDF, polarimetric monitoring of known pulsars and repeating FRB sources has demonstrated that these artifacts in CHIME observations tend to be sub-dominant and usually appear near 0 rad m$^{-2}$.}

The maximum Galactic RM contribution, $\mathrm{RM_{MW}}=50\pm 24\; \mathrm{rad\, m^{-2}}$, is estimated along the sightline of \J{} from the all-sky Faraday Sky map of \citet{Hutschenreuter2022}. This is a map of $\rm{RM_{MW}}$ using RM measurements of 55,190 polarized extragalactic sources. The discrepancy between the observed $\rm{RM}$ and $\rm{RM_{MW}}$ is anomalous, particularly when considering the proximity of \J{} as inferred by its modest DM. There are no obvious structures (e.g., HII, H$\alpha$ regions) that could give rise to large RM contributions. Instead, it seems plausible that a significant fraction of the observed RM is contributed by a magneto-ionic structure near the source, with the structure perhaps relating to the progenitor system. Figure \ref{fig:rm_vs_dm} summarizes this result, showing the \J{} position in DM--RM space relative to the Galactic pulsar sample. Pulsars with comparable $\mathrm{|RM|}$ to \J{} typically have measured DMs that are $\sim10$ times larger than what is measured from \J{}. We note that a separate interpretation where a massive companion is contributing to the local RM is another scenario that could occur, such as in the case of PSR J1259-63 and PSR J1744-24A \citep{Johnston:Wex:Nicastro:2001,Li:Bilous:Ransom:2023}. However, as we see no evidence of binarity in the timing data and we infer a magnetar-like field, we believe that a isolated strongly magnetised source scenario is more plausible. 
\red{Given the proximity of the source, it is somewhat surprising that no associated structure is identified. Using the equation, $B_{\parallel}=1.23\times\frac{\mathrm{|RM|}}{\mathrm{DM}}$ yields a lower-limit of 20 $\rm{\mu G}$. However, the true magnetization of the Faraday active medium may be several orders of magnitude larger given that DM contribution of this medium likely a fraction of the source's total DM.}


  \section{Discussion and Conclusions}
\label{sec:Discussion}
\begin{figure}[ht]
  \centering
  \includegraphics[width=0.8\textwidth]{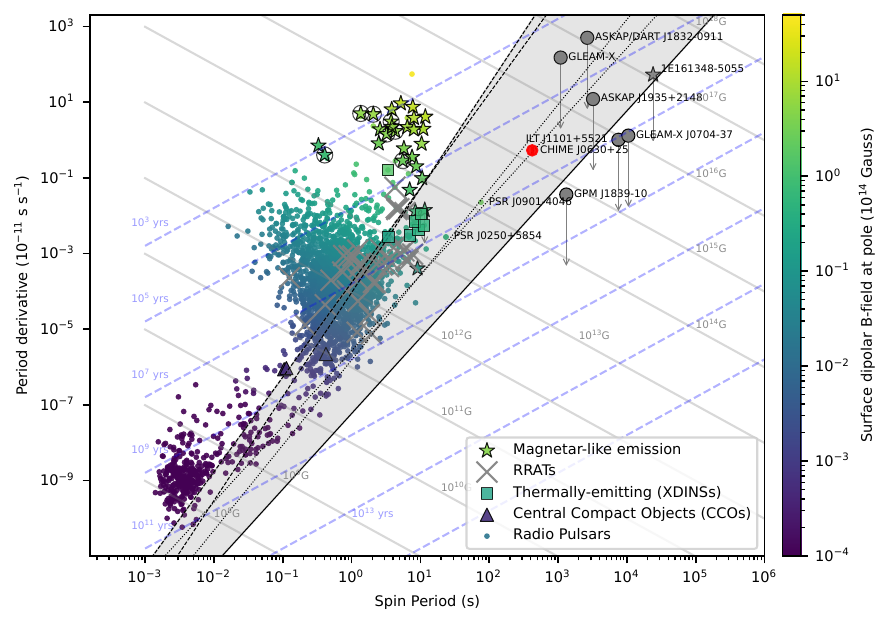}
  \caption{The $P-\dot{P}$ diagram. 1E161348--5055, a central compact object that shows evidence of magnetar-like emission, is shown with a grey star \citep{Dai:Evans:burrows:2016}. The longest period pulsar  PSR J2050+5854 and LPT \meerkat{} are shown by the coloured dots and labelled as such. The grey-shaded region shows the death valley for coherent pulsed radio emission. The black dashed line corresponds to a pure dipole, dotted lines for a twisted dipole, and solid lines for a twisted multipole configuration \citep{kaiyou:malvin:1993,zhang:harding:muslimov:2000}.  The blue dashed lines correspond to lines of constant age, and the grey solid lines correspond to lines of constant magnetic field. All downward-facing arrows correspond to upper limits and are at the 1-$\sigma$ level.}  \label{fig:ppdot}
\end{figure}
Here we discuss the potential sources of \J{}. First, we consider whether the object could be a white dwarf, either isolated or in a binary system. There is no evidence of an abrupt glitch-like spin-up event in a white dwarf to date. In an intermediate polar system -- white dwarf - main sequence binary-- the white dwarf often exhibits a smooth spin-up due to mass accretion \cite[e.g.][]{paice:scaringi:segura:2024}. We do not observe any evidence of a smooth spin up or binarity in \J{}, though it is possible that the large TOA uncertainties and sporadic pulses mask a short-period orbit with a semi-major axis of order one light-second.  A similar type of object, white dwarf pulsars, are white dwarf binaries which are known to emit radio bursts with higher duty cycles and lower luminosities than \J{} \citep[\ARSCO{} and \pelisoliWD{}][]{Marsh:Gansicke:hummerich:2016,Buckley:2017,Pelisoli:2023}. The low duty cycle of \J{}, of 0.4--0.8\%, suggests a tighter beam, like those seen in pulsars. \red{However, this may be complicated by recent discoveries of two LPTs with white dwarfs, \ILT{} and \GLEAM{} with duty cycles between 0.3--2\% \citep{Hurley-Walker:McSweeney:Rea:2024,Rodriguez:2025,deRuiter:Rajwade:Bassa:2025}. In those cases, the radio period and the orbital period of the white dwarf - M-dwarf binary are locked. This seems unlikely in the case of \J{} due to a much shorter period.} Therefore, we conclude that \J{} is unlikely to be a white dwarf.

\red{Comparing the glitch magnitude of \J{} with those known in neutron stars, we find that it is generally consistent with the magnitudes expected. As mentioned in Section \ref{sec:timing}, the fractional glitch amplitude is $|dF/F|=1.3\times10^{-6}$, within the range found in other pulsars from $10^{-9}-10^{-3}$ \citep{McKee:Janssen:Strappers:2016,Serim:Sahiner:Cerri-Serim:2017}. When comparing to magnetars, which have been shown to glitch and anti-glitch (rapid spin-down), we again find general agreement with the glitch amplitudes. For example SGR 1935+2154 was shown to have a spin-down glitch with absolute fractional glitch amplitude of $|dF/F| = 5.8\times 10^{-6}$ \citep{Younes:Baring:Harding:2023}, and 1E 2259+586 was found to have both a spin-up and spin-down of $|dF/F| =1.24 \times10^{-6}$ and $|dF/F| = 5.8\times 10^{-7}$, respectively \citep{Younes:Ray:Baring:2020}. Magnetars are known to show dramatic changes to pulse profile, fluence, and other properties following a glitch. Unfortunately, we have little data with total intensity before the glitch, and we are unable to constrain the changes. We do, however, observe a period of relatively higher activity around and after the glitch epoch (MJD 59341 - 59574).}

Many objects exhibit periodic radio emission at $\BigOSI{1}{h}$, an order of magnitude longer than seen in \J{}. These include flaring ultra-cool dwarfs, like TLV 513-46546 (1.96~h), main sequence stars like CU Virginis (12.5~h) \citep{ravi:hobbs:Wickramasinghe:2010}, and sources of unknown origin like the Galactic Center Radio Transients, GCRT J1745--3009 (1.3h) \citep{Hyman:lazio:kassim:2002} and GCRT J1742--3001 (no period) \citep{hyman:wijnands:lazio:2009}. For ultra-cool dwarfs, it has been suggested that the limit to their spin periods cannot be much less than $\sim$1~h due to rotational break-up \citep{tannock:metchev:heinze:2021}. A similar argument can be given for main sequence stars at about a $\sim$10~h rotational period. GCRTs are pulsing radio sources towards the Milky Way Centre. Like LPTs, their nature remains mysterious. The pulse widths of GCRT J1745--3009 are $\sim$11 minutes wide and much longer than \J{} or any LPT \citep{Hyman:lazio:kassim:2002}. Similarly, GCRT J1742--3001's flares have widths on timescales of months, again much longer than any LPT \citep{hyman:wijnands:lazio:2009}. Therefore, we conclude that none of the known source types with hour-long periods can explain the properties of CHIME J0630+25. 

\red{We have successfully created a phase-connected timing solution \J{} with the detected single pulses. We were able to measure a robust spin-down due to a long timing baseline of $\sim$4.6 years and higher time resolution paired with the relatively narrow pulse-widths of \J{}.}

The timing properties of \J{} strongly suggest a neutron star model. To date, no other periodic coherent burst emitter has exhibited a sudden spin-up of its pulse period apart from neutron stars. Neutron star models for LPTs often invoke an old magnetar. Indeed, \meerkat{} has period of $\sim76$\,s and an inferred magnetar-strength magnetic field of $1.3\times10^{14}$\,G \citep{caleb:heywood:rajwade:2022}. Magnetars are theorized to be the early stage of neutron star evolution, and their strong magnetic fields are predicted to decay on a timescale of $\sim10^{4}$ years \citep{beniamini:hotokezaka:vanderhorst:2019}. From spin-down alone, the maximum period that a magnetar can reach is $\sim13$~s \citep{beniamini:hotokezaka:vanderhorst:2019}. Therefore, it has been theorized that magnetars can only reach longer periods via another mechanism, such as angular momentum kicks via giant flares \citep{beniamini:wadiasingh:metzger:2020}. Such a mechanism would be needed to explain \J{}, given that its inferred magnetic field is $\sim1.5\times10^{15}$\,G, which is into the magnetar regime. We show \J{}'s place among all other neutron stars and comparable sources in Figure \ref{fig:ppdot} \footnote{The code to create this plot was adapted from \citep{Hurley-Walker:Rea:2023}(\url{https://github.com/nhurleywalker/GPMTransient}) with more source categories (i.e., millisecond pulsars and RRATs). \red{The RRATs were retrieved from (\url{https://rratalog.github.io/rratalog/}) and the pulsars from the ATNF Pulsar Catalog V 2.4.0 \citep[\url{https://www.atnf.csiro.au/research/pulsar/psrcat/} ][]{Manchester:Hobbs:Teoh:2005}}}. Furthermore, the burst structure, as seen in 59460A, is reminiscent of known magnetars like XTE 1810-197 \citep{maan:joshi:surnis:2019}. An independent line of evidence from the polarization of \J{} shows that the RM is well above the expected value for this line of sight, and compared to other pulsars, suggesting a strong local magneto-ionic environment and supporting a magnetized neutron star origin.

\red{The majority of our observations resulted in non-detections. We believe that this is due to the intrinsic sporadic nature of \J{}. This is not uncommon amongst pulsars, especially ones found by CHIME/FRB \citep{dong:crowter:meyers:2023}. CHIME/FRB is biased towards finding sporadic sources, because of the large exposure time paired with a single-pulse detection pipeline \citep{10.3847/1538-4357/aad188}. Furthermore, magnetars such as SGR 1935+2154 are known to be sporadic in radio and X-ray \citep{Tavani:Casentini:Ursi:2021,Bochenek:Ravi:Belov:2020,10.1038/s41586-020-2863-y}. With the sporadic nature of \J{} bursts, it is not unexpected that our X-ray observations resulted in non-detections. Our X-ray observations were conducted between MJD 60253 and 60383. Unfortunately, this was far removed from the glitch epoch so that no constraints could be placed on the post-glitch X-ray activity.}

In conclusion, we have discovered a new LPT, \J{}. From timing, we find a glitch and infer that the surface magnetic field is  $\sim 1.5\times 10^{15}$\,G. An independent line of evidence shows that the rotation measure is much higher than expected for this line of sight and for pulsars with similar dispersion measures. We find no evidence of binarity and speculate that \J{} is a highly magnetized, slowly spinning neutron star. Future multiwavelength follow-ups will require arc-second localization, likely achievable by triggered observations with an interferometric radio telescope like the Karl Jansky Very Large Array.

\begin{acknowledgments}
\section{acknowledgements}
We acknowledge that CHIME is located on the traditional, ancestral, and unceded territory of the Syilx/Okanagan people. We are grateful to the staff of the Dominion Radio Astrophysical Observatory, which is operated by the National Research Council of Canada.  CHIME is funded by a grant from the Canada Foundation for Innovation (CFI) 2012 Leading Edge Fund (Project 31170) and by contributions from the provinces of British Columbia, Quebec and Ontario. The CHIME/FRB Project, which enabled development in common with the CHIME/Pulsar instrument, is funded by a grant from the CFI 2015 Innovation Fund (Project 33213) and by contributions from the provinces of British Columbia and Quebec, and by the Dunlap Institute for Astronomy and Astrophysics at the University of Toronto. Additional support was provided by the Canadian Institute for Advanced Research (CIFAR), McGill University and the McGill Space Institute thanks to the Trottier Family Foundation, and the University of British Columbia. The CHIME/Pulsar instrument hardware was funded by NSERC RTI-1 grant EQPEQ 458893-2014.

This research was enabled in part by support provided by the BC Digital Research Infrastructure Group and the Digital Research Alliance of Canada (alliancecan.ca).

The National Radio Astronomy Observatory and Green Bank Observatory are facilities of the U.S. National Science Foundation operated under cooperative agreement by Associated Universities, Inc.

This work made use of data supplied by the UK Swift Science Data
Centre at the University of Leicester and the Swift satellite. Swift,
launched in November 2004, is a NASA mission in partnership with
the Italian Space Agency and the UK Space Agency. Swift is managed
by NASA Goddard. Penn State University controls science and flight
operations from the Mission Operations Center in University Park,
Pennsylvania. Los Alamos National Laboratory provides gamma-ray
imaging analysis.

\allacks
\end{acknowledgments}
  %
  \vspace{5mm}
  \facilities{Swift(XRT and UVOT), CHIME, GBT, uGMRT}


  \software{astropy \citep{astropy:2013, astropy:2018, astropy:2022}, PRESTO \citep{2001PhDT.......123R}, sigpyproc}


  \appendix
\section{Dispersion Measure and Distance Estimate}
\label{dmdist}

The dispersion measure (DM) is a frequency-dependent delay from the ionized component of the interstellar medium affecting all radio pulses. We measured the DM of each pulse using \texttt{DM\_phase} \footnote{\url{https://github.com/danielemichilli/DM\_phase}}, which is a brute force algorithm that maximises the coherent power across the bandwidth by trying many different DMs. Due to the wide burst widths, the DM uncertainty for each pulse is large; this is exemplified in Figure \ref{fig:bursts_1} by the large hot spot that the dedispersion panel covers. 
For the distance determination, the dispersion measure used is the error-weighted average of all the CHIME/Pulsar detections.  
The resultant DM value is 22(1)~pc~cm$^{-3}$. 

With a model of the interstellar electron density, we can use the DM to estimate the distances to the radio source. There exist two models for the Galactic electron density, NE2001 \citep{10.48550/arXiv.astro-ph/0207156} and YMW16 \citep{10.3847/1538-4357/835/1/29}. Analyses of Galactic interstellar medium models have shown that the YMW16 model is better than the NE2001 electron density model \citep{10.48550/arXiv.astro-ph/0207156} for nearby pulsars and is likely accurate to a factor of $\sim$1.5 of the YMW16 estimate \citep{10.1017/pasa.2021.33,10.1088/0004-637X/698/1/250}. Therefore, we used the YMW16 model exclusively. 

The distance was determined by the YMW16 electron density model \citep{10.3847/1538-4357/835/1/29} to be 170\,pc. To find the uncertainties associated with CHIME J0630+25, we queried the ATNF pulsar catalog and applied a series of cuts to the full catalogue. First, we isolated those pulsars with similar Galactic latitudes as the $b=7^{\circ}$ of CHIME J0630+25. We selected pulsars with $b>4^{\circ}$ and $b<10^{\circ}$. We then limited the DM of these sources to be less than 100~pc~cm$^{-3}$. Finally, we only considered pulsars with an independent distance measure, such as parallax or globular cluster association. In total, 13 pulsars met these criteria. To set the uncertainty on the YMW16 value, we found the mean and standard deviation of the ratio between the YMW16 value and the independently derived distance measurements. These are 1.6 and 0.6, respectively. Therefore, we set asymmetric errors about the YMW16 value at 170$^{+310}_{-100}$(95.4\% confidence). We note that this method may be biased low as it is easier to obtain parallax for nearby pulsars.

%



\section{Flux density and spectral index}

\label{calibration}
 Using the total intensity data obtained by CHIME/Pulsar, we measured some essential characteristics such as effective pulse width, fluence, and DM. For the bursts detected from CHIME J0630+25, the fluence is defined by the integrated flux density over the duration of the burst, and the effective width is calculated via $W_{\rm eff}=\text{F}/\text{S}_{peak}$ where F is the fluence and S$_{peak}$ is the peak flux density. 
CHIME/Pulsar has declination-dependent sensitivity. Therefore, we used 3C133 as a calibrator source to determine the system equivalent flux density (SEFD). The SEFD is calibrated by fitting the telescope temperature and is defined in the following way
\begin{equation}
  \text{SEFD}(\nu) = \frac{T_{\rm telescope}(\nu)+T_{\rm sky}(\nu)}{G(\nu)}\label{eq:sefd}
\end{equation}
where $T_{\rm sky}$ is obtained from the Haslam 408-MHz all-sky map \citep{10.1093/mnras/stv1274,1982A&AS...47....1H}, $G$ is the telescope gain and $T_{\rm telescope}$ is the system temperature of all telescope components (i.e. receiver, structure, ground etc). The flux density of a source is given by \begin{equation}
  S(\nu) = \frac{T_{\rm on}(\nu)-T_{\rm off}(\nu)}{T_{\rm off}(\nu)}\times \text{SEFD}
  (\nu),
\end{equation}
where $S(\nu)$ is the flux density. $T_{\rm on}$ and $T_{\rm off}$ are the temperatures of the calibrator source and a blank patch of nearby sky, respectively. $T_{\rm telescope}$ is an unknown. Thus, we performed a maximum likelihood reduced $\chi^{2}$ fit of the CHIME/Pulsar measured 3C133 spectrum against the catalogued flux density measurements of 3C133. We used the VLA calibrator list \footnote{https://science.nrao.edu/facilities/vla/observing/callist} for the catalog values. We found that the best functional form of $T_{\rm telescope}$ is a 5th-order polynomial. Refer to \cite{Dong_2024} for more details. 

Using the total intensity data for the CHIME/Pulsar bursts, we measured the spectral index of CHIME J0630+25. First, we calibrated the flux density of the bursts using the method described above. This process will also serve to calibrate the spectrum of CHIME J0630+25. Then, each burst is integrated over its duration to obtain a spectrum. Maximum likelihood is used to fit the spectra with a power law model of the form
\begin{equation}
  S(\nu) = A\nu^{\alpha}
\end{equation}
where $\alpha$ is the spectral index, and A is the amplitude parameter. Both $\alpha$ and A are fit parameters. The spectral indices are provided in Table \ref{tab:burst_characteristics_1}.

To place an uncertainty on the fit, we measured the spectral index of 23 other calibrator sources with known spectral indices using the same technique. We found that the mean uncertainty on the spectral index calibrated in this way is 0.3. 


\section{Follow up of \J{}} \label{sec:followup}
\label{observations}
\begin{figure}[ht]
  \centering
  \includegraphics[width=\textwidth]{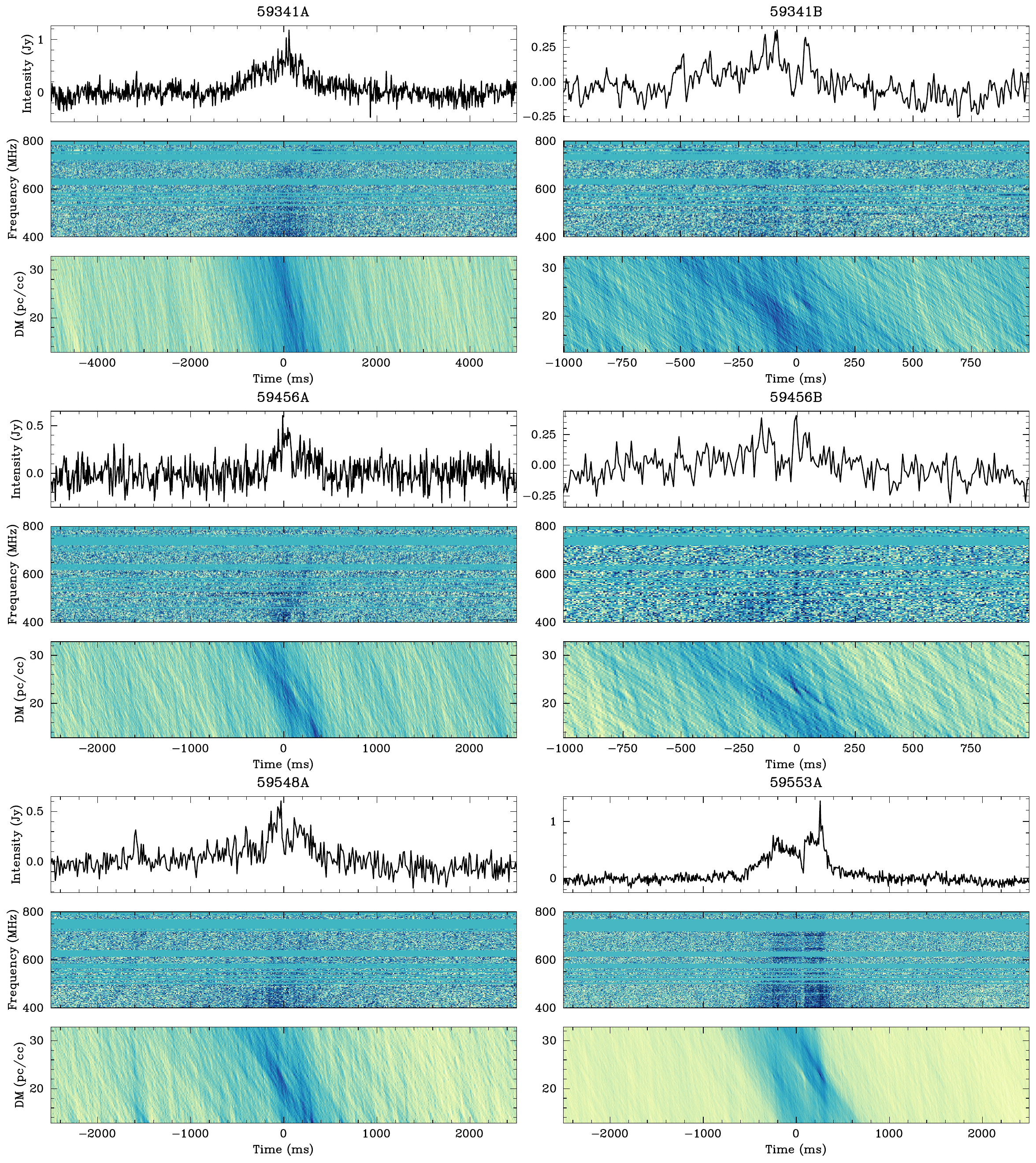}
  \caption{Additional dynamic spectra analogous to Figure \ref{fig:bursts_1}.}
  \label{fig:bursts_2}
\end{figure}
\begin{figure}[ht]
  \centering
  \includegraphics[width=\textwidth]{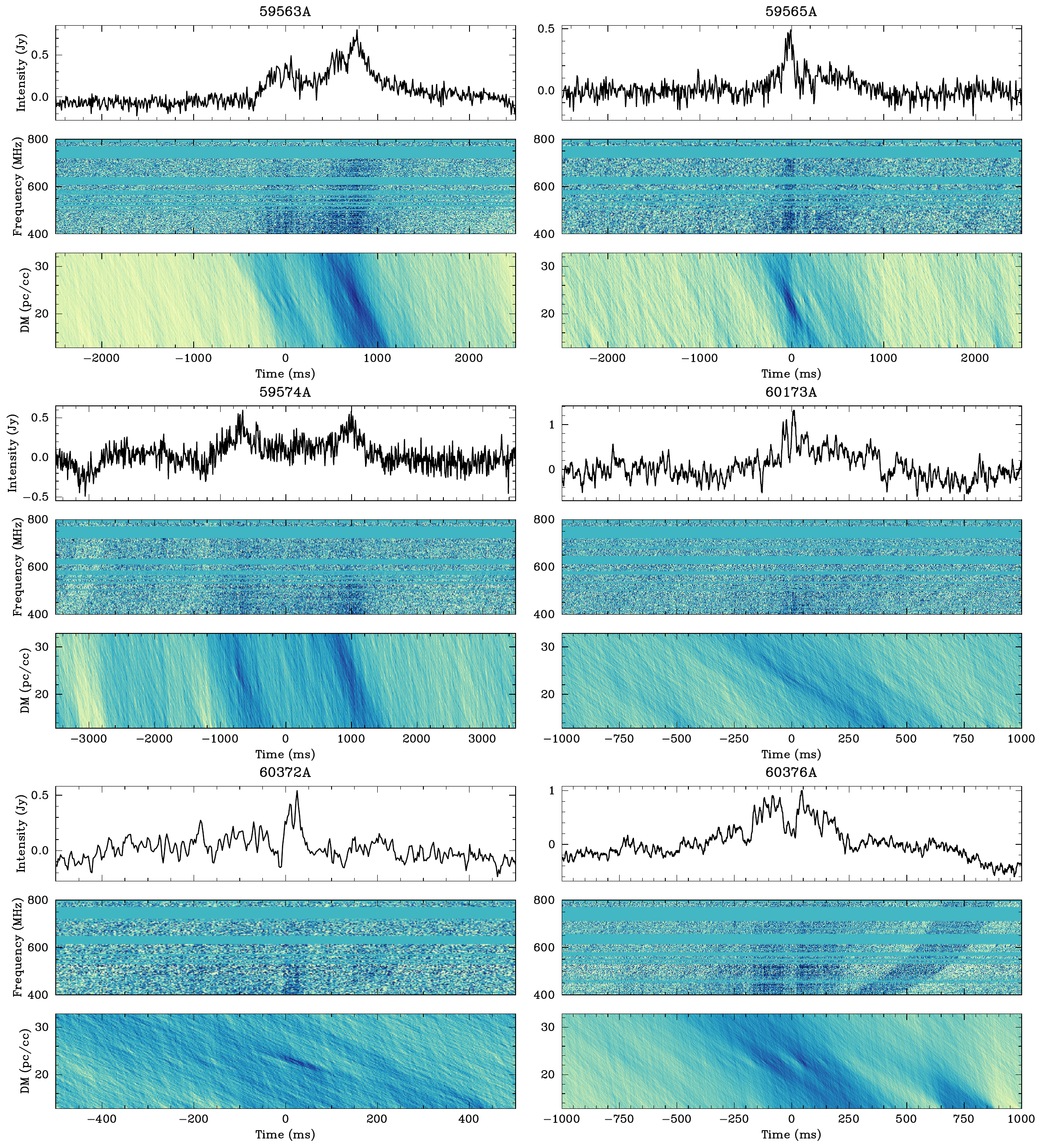}
  \caption{Continued.}
  \label{fig:bursts_3}
\end{figure}
\begin{figure}
  \centering
  \includegraphics[width=0.5\textwidth]{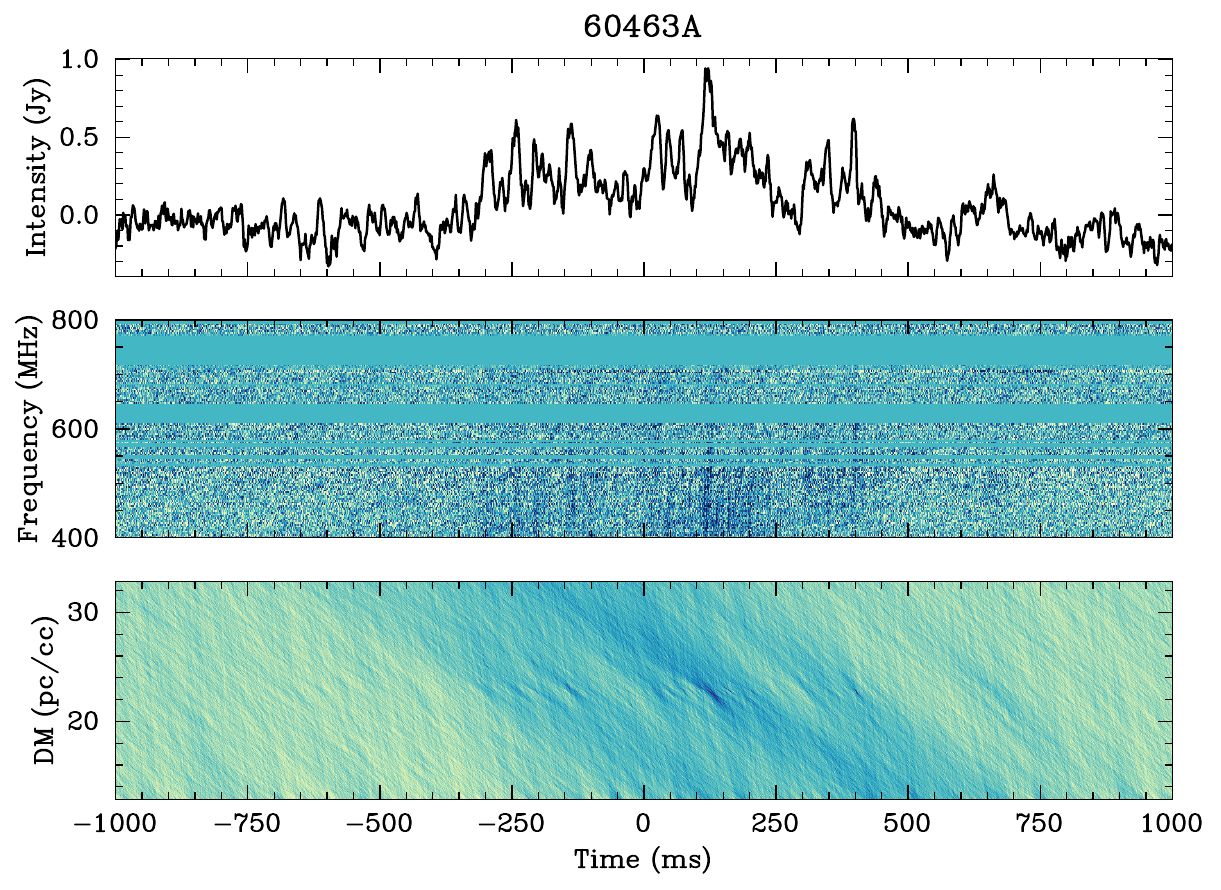}
  \caption{Continued.}
  \label{fig:bursts_4}
\end{figure}

To process all the CHIME/Pulsar data, we have employed the CHIME/Pulsar Single-pulse PIPEline (CHIPSPIPE)\footnote{https://github.com/CHIME-Pulsar-Timing/CHIME-Pulsar\_automated\_filterbank}, an automated single-pulse search pipeline designed to handle the large data volume of CHIME/Pulsar. The pipeline is based on PRESTO \citep{2001PhDT.......123R}. It further utilises SPEGID \citep{pang:Goseva-Popstojanova:devine:2014} and FETCH \citep{agarwal:aggarwal:burke-spolaor:2020} to filter out spurious candidates. Finally, the pulses that are graded as astrophysical by FETCH are examined by a human. For a detailed discussion of CHIPSPIPE, refer to \cite{dong:crowter:meyers:2023}. For all the bursts that passed the human check, we manually removed the RFI-contaminated channels and generated the dynamic spectrum, pulse profile, and dedispersion-time plot for each burst using the \texttt{Your} \citep{10.21105/joss.02750} package. 
CHIPSPIPE has limited sensitivity to wider pulses. Therefore, we visually inspected the dynamic spectrum and pulse profile around each detected pulse within a 10-second window. This manual inspection aims to identify additional sub-pulses missed by CHIPSPIPE that may provide evidence of quasiperiodicity. 59460A, 59463A, 59553A, 59563A, and 59574A exhibited distinct second peaks not flagged by the initial CHIPSPIPE detection.

We use data from both archival observations near \J{} and targeted follow-up campaigns to explore \J{} across many wavelengths. In the radio band, we used the Green Bank Telescope (GBT) because of its increased sensitivity and longer observation tracks compared to CHIME. We also used the upgraded Giant Metrewave Radio Telescope (uGMRT) and archival VLA Low-band Ionosphere and Transient Experiment (VLITE). Pulsars and magnetars are known to emit X-rays. Therefore, we also performed targetted observations with the Neils Gehrels Swift Observatory's X-Ray Telescope (XRT). Some magnetars are soft $\gamma$-ray repeaters. Thus, we also searched the known $\gamma$-ray archives for as-of-yet-unknown magnetar candidates. 
\red{As new observations have also revealed some LPTs are binary systems with optical counterparts \citep{deRuiter:2024, Hurley-Walker:2024}, we search through archival optical data for a possible time-domain optical counterpart to \J{} as well.}

With the GBT, we used the VEGAS back end at 20.48~$\mu$s time resolution and coherently dedispersed to 22.5~pc~cm$^{-3}$. The observations were taken between MJD 59861 and 59914. The Stokes I data were processed using CHIPSPIPE to search for single pulses. No pulses were detected. \redtwo{With the single pulse radiometer equation,
\begin{equation}
    S = \frac{\textrm{S/N}\, S_{sys}\beta}{\eta\sqrt{n_p W \Delta \nu}}
\end{equation}
we are able to estimate the minumum fluence. Where $S$ is the flux, S/N=6 is the minimum signal to noise, $S_{sys}\approx15$\,Jy is the system equivalent flux density including the Galactic background, $\beta=1$ is the telescope degredation factor, $\eta=0.868$ is a Gaussian pulse shape approximation, $n_p=2$ is the number of polarizations, $W=2$\,s is an estimate of the pulse width, and $\Delta \nu=0.24$\,GHz is the bandwidth of the 800\,MHz Prime Focus receiver. The derived minimum fluence for a 2\,s burst is $\sim$7\,Jy\,ms. We caution, however, that the actual value may be higher. Our detection pipeline preferentially detects bursts which are ``spikey'' with an easily detectable DM-hotspot. The quantification of this is difficult and will require an intensive campaign of synthetic burst injection and retrieval, which is out of the scope of this study.}
With the uGMRT, we performed the observations between MJD 59700 and 59837 from 950 MHz to 1460 MHz. These had an integration time of 0.67 s in incoherent array mode. Unfortunately, due to the variable baseline and the high levels of radio frequency interference (RFI), we were not able to make use of the uGMRT data. Finally, we searched through archival data from VLITE \citep{clarke:wendy:walter:2018}\footnote{\url{https://vlite.nrao.edu}} for high-resolution observations ($\sim$ 5") at 340~MHz covering the region of interest. We identified two observations where \J{} was located within 2$^{\circ}$ of the phase center and made short time interval images at the VLITE sample time of 2~s. The short VLITE images were catalogued using PyBDSF \footnote{\url{https://github.com/lofar-astron/PyBDSF}}, and components associated with all persistent radio sources were eliminated. The remaining catalogued sources were low signal-to-noise (S/N $\sim$ 4), and visual inspection revealed these remaining candidates were likely associated with poorly cleaned sidelobes in the images. As the CHIME/FRB localisation is large compared to the VLITE resolution, the chance coincidence for a low S/N candidate is large. Therefore, it is difficult to associate any low S/N VLITE candidates with \J{}. Unfortunately, VLITE's highest time resolution is 2~s, and so bursts from \J{} are predominantly less than one bin long, resulting in low S/N. This makes it challenging to differentiate potential \J{} bursts in VLITE data from remaining uncleaned sidelobe structures. 

Our X-ray observations of \J{} consisted of 32ks of Swift XRT time under target ID 97140 and 97203. To process the data, we used the tools provided by the UK Swift Science Data Center \footnote{https://www.swift.ac.uk/user\_objects/} to create the images. Then, we used \texttt{Ximage} \footnote{https://heasarc.gsfc.nasa.gov/docs/software.html} to detect the sources and provide S/N estimates. No sources were detected within the 3$\sigma$ error region of \J{}. Therefore, we use the Swift XRT data to place a flux limit on the source location. With a background count rate of 4.6$\times$10$^{-4}$ counts/s, we place a $3\sigma$ unabsorbed flux upper limit of 1.2$\times$10$^{-14}$ erg cm$^{-2}$ s$^{-1}$ and 3.2$\times$10$^{-14}$ erg cm$^{-2}$ s$^{-1}$ for a blackbody and power-law spectrum, respectively. \redtwo{This corresponds to $4-16\times10^{28}$\,ergs\,s$^{-2}$. If, however, one assumes the 95\% upper limit on the distance, 520\,pc, then the upper limits are $3.9-15\times10^{29}$\,ergs\,s$^{-1}$}. The blackbody spectrum is assumed to have a temperature of 0.3 keV, and the power-law spectrum is assumed to have a photon index of 1.0. The flux limits are calculated using \texttt{WebPIMMS}\footnote{\url{https://heasarc.gsfc.nasa.gov/cgi-bin/Tools/w3pimms/w3pimms.pl}}. \redtwo{Comparing the median limits to other objects from Figure 10 in \cite{10.1093/mnras/staf1227}, we find that the X-ray luminosity constraint is $\sim1-2$ orders of magnitude lower than the next brightest LPT, ASKAP J1935+2148 \citep{caleb:lenc:kaplan:2024}, and ASKAP J1448-6856 \citep{10.1093/mnras/staf1227}. Furthermore, \J{} is $\sim1$ order of magnitude lower than the least luminous X-ray bright pulsar. However, many pulsars are not detectable in X-rays.}

We also searched for possible as-of-yet unknown soft $\gamma$-ray repeater counterparts to \J{}. These would reside in the same databases as $\gamma$-ray bursts (GRB), albeit with an unknown classification. We first cross-match the coordinates and times of arrivals of \J{} with all $\gamma$-ray sources reported in GRBWeb \footnote{\url{https://user-web.icecube.wisc.edu/~grbweb_public/}}. We limit the GRBWeb triggers to those that are well localised (e.g., 1$\sigma$ spatial error $<$1 degree), as it is challenging to claim significant spatial coincidences for triggers with either unknown or large uncertainty regions. In our cross-match, we conservatively assume a 1$\sigma$ positional error in RA of 1 degree and a 1$\sigma$ positional error in DEC of 0.5 degrees for \J{}. We then cross-match the localisation region of \J{} with that of all known sources in GRBWeb, requiring the localisations to be consistent within the 3$\sigma$ uncertainties. Within one week of each burst, we did not find any sources to be coincident with \J{}. However, given GRBWeb's focus on cosmological GRBs and not Galactic $\gamma$-ray sources such as soft $\gamma$ repeaters, we also cross-match the position of \J{} and its bursts with all triggers reported in the $\gamma$-ray Coordination Network (GCN)\footnote{\url{www.gcn.gsfc.nasa.gov}} circulars. We again limit our search to well-localised triggers, e.g., ($\sigma<1$ degree), and do not find any trigger-burst pairs with the given criteria. When considering solely spatial coincidence, however, we find one trigger spatially coincident with \J{}. The trigger is GRB110414A, which was detected long before CHIME was built.

However, as noted in \cite{curtin:tendulkar:josephy:2023}, there is a high chance probability of finding spatial coincidences given CHIME's current localization capabilities. Accordingly, we conclude that no significant coincidences exist between \J{} and any known $\gamma$-ray triggers. 

\red{
The two known optical counterparts to LPTs (ILT\,J1101+5521, GLEAM-X\,J0704$-$37) are detached white dwarf + M dwarf binary systems, establishing that magnetic white dwarfs are linked to at least some LPTs.
In those systems, the radio pulsations are phase-aligned with the orbital period \citep{deRuiter:2024, Hurley-Walker:2024, Rodriguez:2025}.
While for \J{}, $P \sim 7$~min is far too short an orbital period for a binary system involving a main sequence star, such a period can be indicative of other possibilities, e.g., the radio pulsations reflecting the spin-period of a white dwarf in an intermediate polar system.
Thus, even though we believe \J{} is unlikely to be a white dwarf (Section~\ref{sec:Discussion}), we search for a time-domain optical counterpart to \J{}.
}

\red{
To conduct this search, we use results from a bulk periodicity search \citep{2020ApJ...905...32B} on data from the Zwicky Transient Facility \citep[ZTF;][]{2019PASP..131a8002B, 2019PASP..131g8001G}.
We restrict the results of the search for periodic time-variable sources within a conservative 5$\sigma$ localization uncertainty region of \J{}. We also restrict our search to be within a period $421 \pm 120$~s in order to encompass the possibility of slightly misaligned optical and radio frequencies \citep[e.g., beat frequencies, as is observed in intermediate polars and white dwarf pulsars;][]{2002A&A...384..195N, 10.1038/nature18620, Pelisoli:2023}. We find $\sim$150 marginal sources exhibiting periodicity at marginal significance, 2 of which have a blindly identified period within $\sim$1~s of that of \J{}. However, visual identification of the light curves of the sources folded to their identified periods show the data are noisy, and that most of these marginal candidates (including the two candidates with close identified periods) are likely false positives.
}

\red{Due to the relatively low galactic latitude ($|b|\sim7^\circ$), we note that many optical sources are visible in the localization region of \J{}.
There is also $E(B-V) \approx 0.3$~mag of dust extinction towards \J{} that can obscure fainter optical sources, for which infrared observations may reveal even more sources. While spectrophotometric follow-up of the identified marginal candidates can confirm or refute the significance of the identified periodicities, the current density of sources in the localization region will make it difficult to robustly identify an optical counterpart to \J{} without a more refined localization.}
\section{F0-F1 grid search}\label{sec:appen_timing}
\begin{figure}
  \centering
  \includegraphics[width=0.5\textwidth]{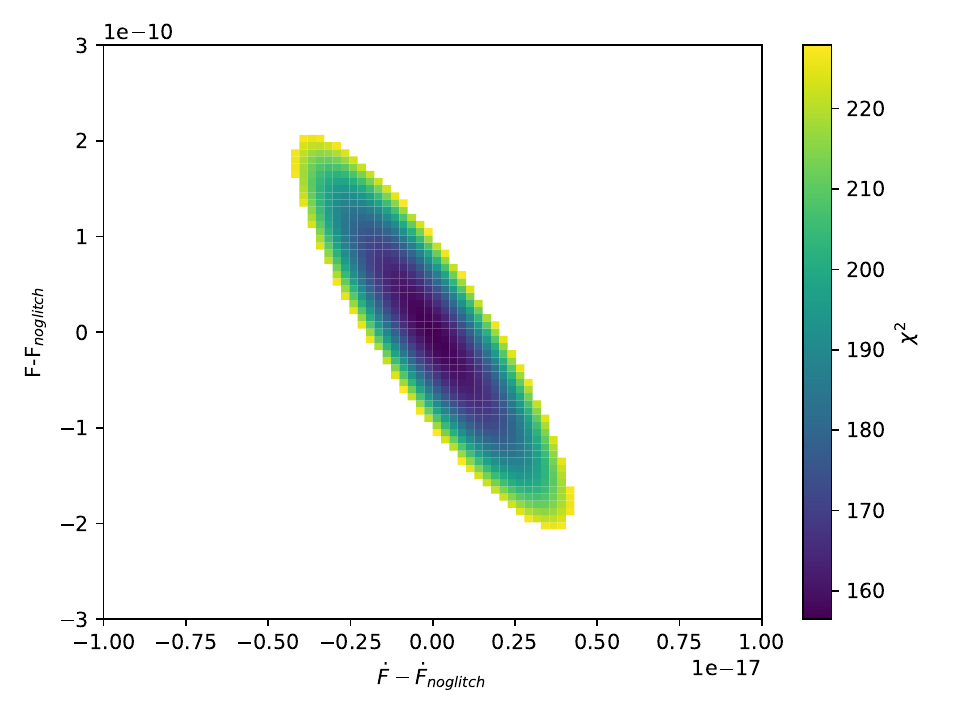}
  \caption{We show the only minimum when we perform a grid search over frequency and frequency derivative parameters. The $\chi^2$ increases rapidly as we deviate away from the optimum parameters; therefore, we choose to zoom in on the minimum.}
  \label{fig:F0_F1_grid}
\end{figure}

\red{To test whether the non-glitch solution converged on a local minimum via PINT's downhill weighted least squares algorithm, we performed a full grid search over frequency (F0) and frequency derivative (F1) parameters, keeping RA and Dec fixed. We searched over a large parameter range from $-3\times10^{-9}<F-F_{noglitch}<3\times10^{9}$ and $-10^{-15}<\dot{F}<10^{-15}$ with 10,000 trials in the grid. We find no other minimums apart from the one identified by our timing solution in Table \ref{tab:J0630+25_timing}. We show the $\chi^{2}$ contour of the minimum in Figure \ref{fig:F0_F1_grid}.}

\section{Metadata Time of Arrival Test}\label{sec:appen_J0012}
\begin{figure}
  \centering
  \includegraphics[width=0.49\textwidth]{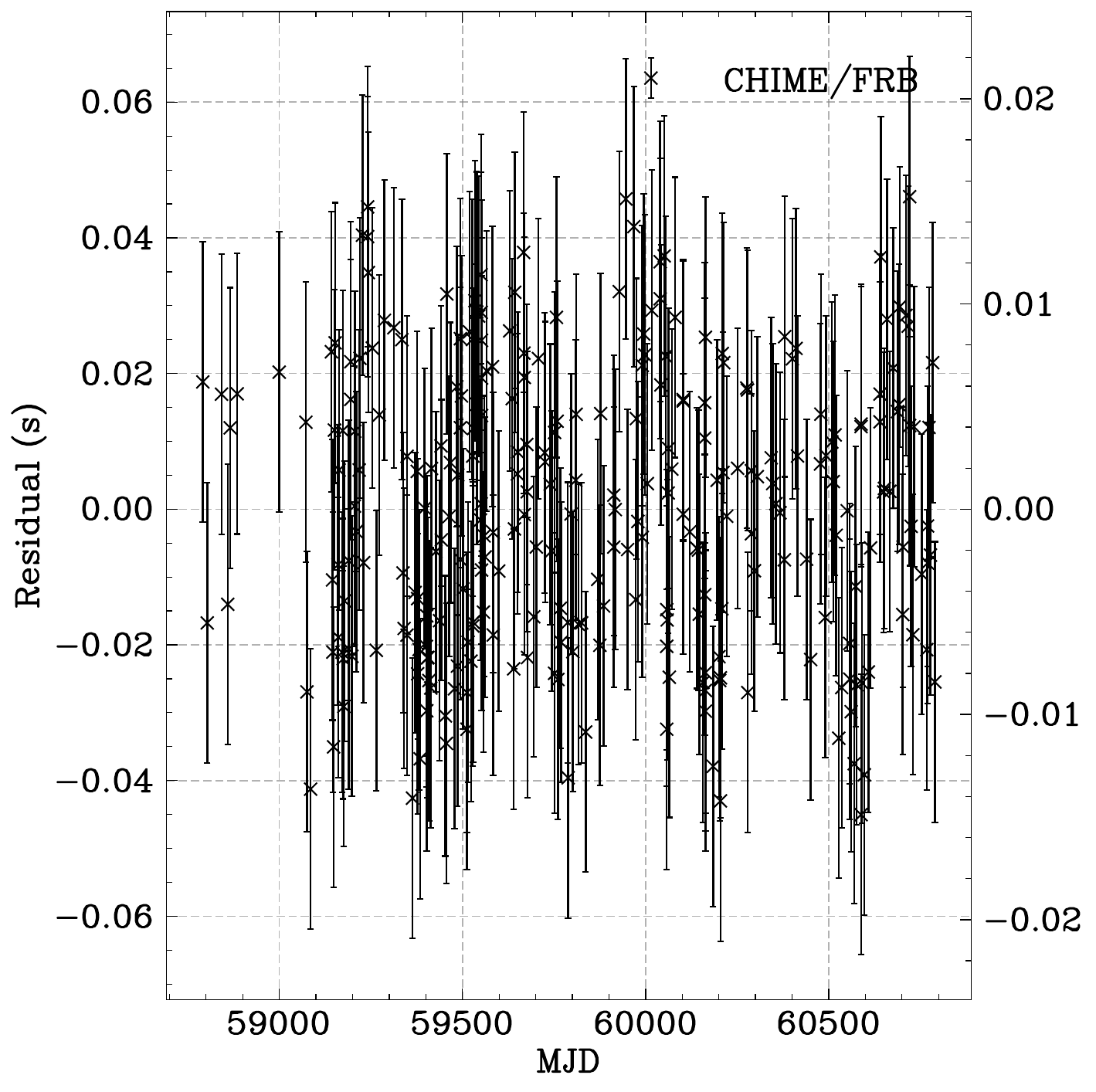}
  \includegraphics[width=0.49\textwidth]{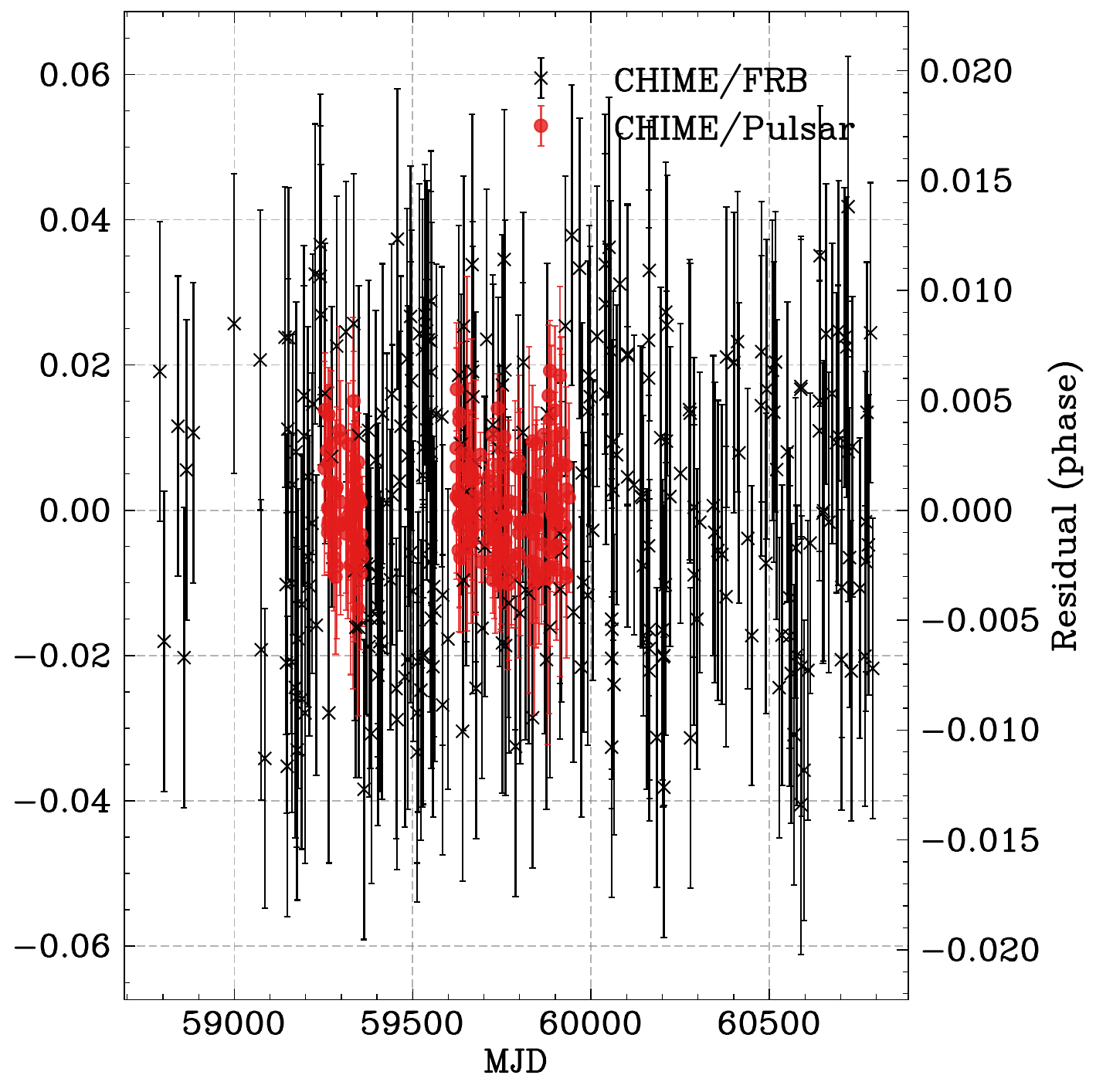}
  \caption{The timing residuals for PSR J0012+5431. On the left we show timing residuals fit with only using CHIME/FRB metadata, on the right we show a combination of CHIME/Pulsar intensity data and CHIME/FRB metadata.}
  \label{fig:0012_resids}
\end{figure}
\red{We test the robustness of the TOAs derived from metadata only by timing a bright ``regular'' pulsar, PSR J0012+5431, via single pulses. PSR J0012+5431 is a rotating radio transient and shows stable timing properties when timed via single pulses in \cite{dong:crowter:meyers:2023}. We show the residuals in Figure \ref{fig:0012_resids}. We start with the timing solution presented in \cite{dong:crowter:meyers:2023} and determined that $\sim20$\,ms errors on each TOA is sufficient to produce a $\chi^{2}_{reduced}=1$. Therefore, the upper limit on errors of each metadata TOA is $\sim20$\,ms as the noise budget of the timing of PSR J0012+5431 is a combination of the TOA error and pulse-to-pulse jitter. This is much lower than the 1\,s errors which we assign to \J{} metadata detections. Furthermore, to explore if there are any systematic/secular deviations between the metadata and the CHIME/Pulsar intensity data we perform a combined fit between the two dataset with a jump as discussed in Section \ref{sec:timing}. We do not find any evidence of any systematic/secular deviations. From this we conclude three things, our assumption that TOA errors are dominated by the pulse shape is valid, the metadata TOAs are sufficiently accurate for the timing of \J{}, and the combination of metadata and intensity data shows no systematics apart from the known jump.}

  \bibliography{J0630+25.bib}
  \bibliographystyle{aasjournal}
\end{document}